\title{Discovery of an explicit closure dispersion model for contrast-agent transport in arteries}
\newcommand{\AuthorA}{Yi-Teng Wang}
\newcommand{\AuthorB}{Yu-Xing Gong}
\newcommand{\AuthorC}{Yan Qiu}
\newcommand{\AuthorCorrA}{Kai-Rong Qin}

\newcommand{\Om}[1]{10^{#1}}
\newcommand{\avg}[1]{\langle{#1}\rangle}
\newcommand{\iunit}{\vmathbb{i}}
\newcommand{\e}{\mathrm{e}}
\newcommand{\Fourier}[1]{\mathcal{F}\left[{#1}\right]}
\newcommand{\F}[1]{\hat{#1}}

\documentclass[lineno]{JFM-FLM_Au}

\usepackage{cleveref}
\usepackage{subcaption}
\crefname{figure}{Figure}{Figures}
\crefname{table}{Table}{Tables}
\crefname{section}{Section}{Sections}
\crefname{equation}{Equation}{Equations}
\crefrangeformat{equation}{Equations~(#1-#2)}

\lefttitle{Yi-Teng Wang et al.}
\righttitle{Journal of Fluid Mechanics}

\author{\AuthorA \aff{1}, \AuthorB \aff{2}, \AuthorC \aff{1}, \and \AuthorCorrA \aff{1}}

\affiliation{\aff{1}School of Biomedical Engineering, Faculty of Medicine, Dalian University of Technology, Dalian 116024, P. R. China, \aff{2}College of Biomedical Engineering, Fudan University, Shanghai 200433, P. R. China}

\corresau{\AuthorCorrA, krqin@dlut.edu.cn}

\begin{document}
\maketitle
	
\begin{abstract}
At high P\'eclet number, the classical Taylor--Aris dispersion model becomes inadequate for early-time contrast-agent transport in arteries, while an explicit closure model and a clear physical interpretation of this regime remain lacking. In this study, we develop a novel explicit-closure one-dimensional (1-D) effective dispersion model for this regime, with its functional structures identified through symbolic regression. Analysis of the resulting model reveals that, in the high-$\Pen_r$ regime, axial transport is redistributed between the effective convection flux and the dispersive flux, resulting in a reduction of the effective convective transport velocity in the dispersion model. This redistribution gives rise to a transition from the classical quadratic scaling to a linear scaling of the effective diffusivity $D_{\mathrm{eff}}$ with radial P\'eclet number $\Pen_r$. Numerical validation demonstrates close agreement with the convection-diffusion model over the investigated high-$\Pen_r$ conditions, while the classical Taylor--Aris model exhibits substantial deviations. Application of the proposed model to averaged flow velocity inversion further demonstrates improved velocity estimation, particularly in the high-$\Pen_r$ regime. These results highlight the importance of accounting for non-classical dispersion for reliable contrast-agent-based arterial blood flow velocimetry and provide new insight into high-$\Pen_r$ mass transport.

\end{abstract}
	

\section{Introduction}
\label{sec:intro}

One-dimensional (1-D) effective dispersion models, particularly those with explicit closures, provide tractable representations of the macroscopic transport behaviour of contrast agents in pulsatile arterial blood flow while avoiding the computational cost of fully three-dimensional (3-D) convection-diffusion models \citep{FimbresWeihs2010, Oldrini2021, Karmakar2022}. Such models are also central to blood-flow velocimetry based on medical imaging \citep{Shpilfoygel2000, WaechterStehle2012, Huang2013, Wu2018, Hoffman2021}. Reliable explicit-closure dispersion models are therefore important for both understanding macroscopic transport mechanisms and inferring flow velocity from medical images.

The theoretical foundation of 1-D dispersion models in laminar flows was established by G. I. Taylor and Rutherford Aris \citep{Taylor1953, Aris1956, Aris1960}, and is now known as Taylor--Aris dispersion. Since then, extensive efforts have been devoted to the development and refinement of 1-D dispersion models for steady laminar flows in rigid conduits \citep{Gill1970, Gill1971, Gill1972, Chatwin1973, Chatwin1975, Smith1982, Yasuda1984}. More recently, these models have been further examined in quasi-steady regimes \citep{Qin2010, Zhu2013} and extended to pulsatile flows \citep{Wang2025}.
Despite the progress above, these models rely on a key assumption that transverse molecular diffusion is sufficiently rapid in contrast of the axial convection to homogenize the solute concentration across the cross-section. This requirement is typically expressed in terms of a small radial P\'eclet number.
Consequently, classical analytical approaches to dispersion often rely on slow-manifold reductions or long-time asymptotics \citep{Mercer1990,Marbach2019,Ding2024,Ding2025}. Within these frameworks, both numerical \citep{Wang2025, Wu2022, Pal2025} and mathematical \citep{Mikelic2011, Berg2020, Sarkar2026} studies based on the 1-D dispersion models mainly focus on the regime where radial P\'eclet number is small enough to prevent violating the assumptions above.

\begin{figure}
	\centering
	
	\begin{minipage}{0.95\textwidth}
		\makebox[\linewidth][l]{(a)}
		\centering
		\includegraphics[width=\linewidth]{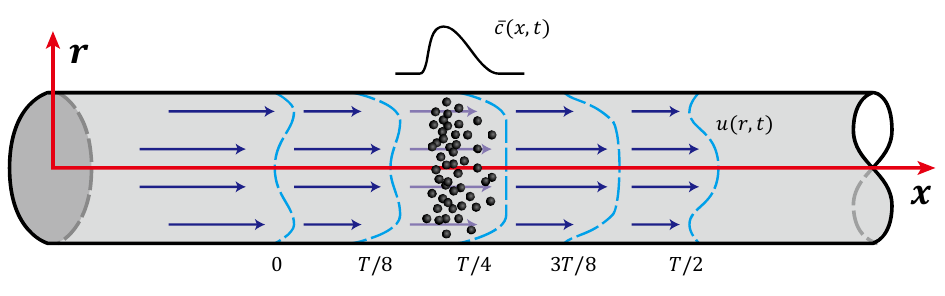}
	\end{minipage}
	
	\begin{minipage}{0.95\textwidth}
		\makebox[\linewidth][l]{(b)}
		\centering
		\includegraphics[width=\linewidth]{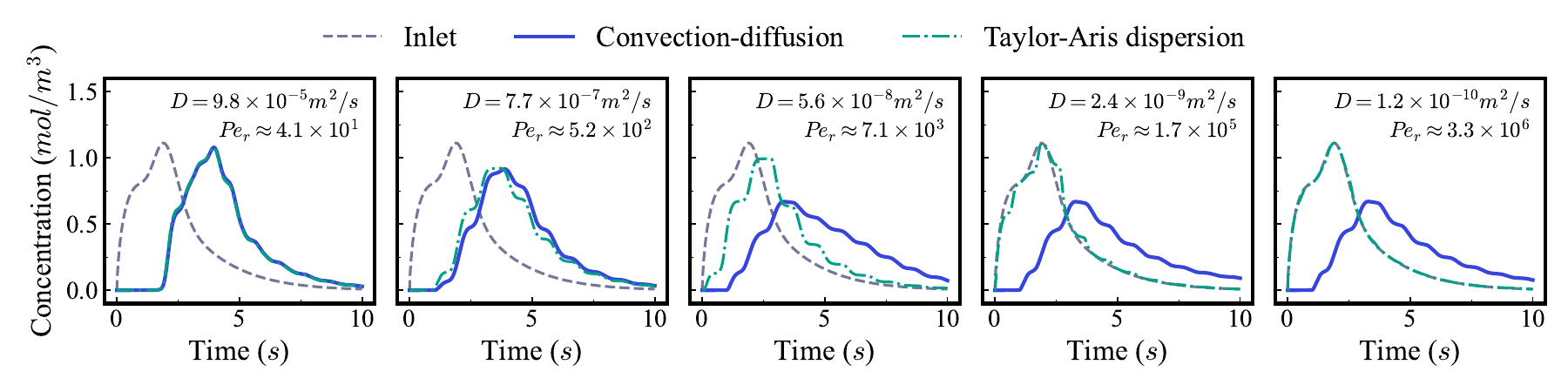}
	\end{minipage}
	
	\caption{(a). Illustration of the solute transport under pulsatile flow velocity profile in a long slender rigid tube. The blood flow velocity profiles of different phases during one cardiac cycle in cylindrical coordinate system are illustrated, which known as the Womersley flow velocity profile \citep{Womersley1955}. (b). Comparison of the numerical solution of the convection-diffusion and Taylor--Aris dispersion equation under different cases. The physical parameters are set as follow: characteristic flow velocity $U=200 mm/s$, tube length $L=400 mm$, tube radius $R=2 mm$, diffusion coefficient $D$ is set from $9.8\times10^{-6} m^2/s$ to $1.2\times10^{-10} m^2/s$, the radius P\'eclet number $\Pen_r$ increases from $\Om{1}$ to $\Om{6}$ correspondingly. As the radial P\'eclet number increases, the shear-induced enhancement of axial dispersion predicted by the classical Taylor--Aris theory is increasingly overestimated, leading to a significant deviation between the Taylor--Aris predictions and the numerical results of the full convection-diffusion equation. In particular, for $\Pen_r \ge \Om{5}$, the Taylor--Aris predictions closely overlap with the inlet concentration boundaries, indicating the breakdown of the conventional Taylor--Aris framework in this regime.}
	\label{fig:problem}
\end{figure}

However, in the transport of contrast agents in arterial flows, the diffusivity of the contrast agent is on the order of $\Om{-10} m^2/s$ \citep{Gillis2000}. Combined with the high flow velocities encountered in arteries, this leads to radial P\'eclet numbers within the high-P\'eclet-number range considered in this study, which is physiologically motivated.
This indicates that the diffusion of the contrast agent in radial direction is always insufficient during the transport. As a result, the fundamental assumption underlying classical Taylor--Aris theory breaks down. Under such high-P\'eclet-number conditions, the shear-induced enhancement of axial dispersion predicted by the classical theory is significantly overestimated, leading to substantial errors in the predicted axial transport and, ultimately, the loss of validity of the conventional Taylor--Aris framework in this regime, as it's illustrated in \cref{fig:problem} (b).

For contrast-agent transport in arterial flows, we identify three key requirements for a reliable dispersion model: (i) the high-P\'eclet-number regime relevant to arterial transport; (ii) the early-time regime, in which the solute concentration remains radially non-uniform; and (iii) an explicit 1-D closure that does not require resolving the underlying transverse transport problem.
Existing studies have addressed these requirements only partially. Early-time dispersion has been investigated at relatively low P\'eclet numbers \citep{Berg2020, Taghizadeh2020}, whereas analytical scaling relations for high-P\'eclet-number dispersion have primarily been derived for the long-time limit \citep{Koch1985}. Convolution-based formulations can describe dispersion over a broad P\'eclet-number range \citep{Wang1983}, but still require the underlying transverse transport problem to be solved and therefore do not provide an explicit 1-D closure.
Consequently, a reliable 1-D dispersion model with an explicit closure for early-time, high-P\'eclet-number contrast-agent transport in arterial flows is still lacking, and the underlying transport mechanisms in this regime have yet to be clearly elucidated.

In another aspect, a substantial body of work has investigated longitudinal solute transport under high-P\'eclet-number ($\Pen_r > \Om{3}$) conditions directly through 3-D numerical simulations or experiments. Despite the limitations of classical Taylor--Aris theory, both numerical \citep{Pereira2005, Ortan2009, Adrover2013, Gekle2017, Wang2014, Adrover2019} and experimental \citep{Bijeljic2004, Pereira2005, Tsakiroglou2005, Boschan2008, Zhang2019} studies have characterized dispersion properties using approaches such as moment analysis. However, no explicit closure model for longitudinal dispersion has been established from these results. These studies indicate that, even in regimes where classical Taylor--Aris theory breaks down, Taylor-dispersion-like behaviour can still be observed, while its underlying parametric structure becomes difficult to characterize using conventional analytical approaches. In particular, the breakdown of cross-sectional homogenization undermines the regular perturbation and asymptotic-expansion frameworks that would otherwise enable the systematic derivation of an explicit dispersion closure.

Nevertheless, the failure of these classical derivation approaches does not imply the absence of an effective 1-D closure in this regime. This motivates the use of symbolic regression to identify the functional structure of the 1-D closure directly from the transport dynamics. Conventional symbolic regression typically relies on a predefined library of mathematical operators, which can make the search increasingly demanding for nonlinear mappings involving multiple parameters. The Kolmogorov--Arnold network (KAN), introduced by \cite{Liu2024}, offers an alternative by learning univariate functional mappings directly from data, allowing the functional structure to emerge without prescribing specific forms a priori and with relatively low computational cost \citep{Zhang2025, Khedr2025, Ouyang2026, Du2026}. This provides a practical approach for identifying the closure structure and subsequently interpreting the underlying transport mechanism.

In this study, we investigate the breakdown of classical Taylor--Aris dispersion in the high-P\'eclet-number regime and develop an explicit 1-D closure for early-time contrast-agent transport in arterial flows. We first examine the transport dynamics underlying the breakdown of cross-sectional homogenization and then identify the functional structure of the closure directly from full convection-diffusion simulations using transfer-function matching and a modified KAN. The proposed model is subsequently examined from both forward- and inverse-problem perspectives, with its predictive performance and physical implications assessed against the full convection-diffusion dynamics.

The main contributions of this study are threefold. First, we establish an explicit 1-D dispersion closure for early-time transport at high P\'eclet numbers, a regime not addressed by existing explicit closure models. Second, the proposed model reveals a transition from the classical quadratic scaling to a linear scaling of the effective diffusivity $D_{\mathrm{eff}}$ with radial P\'eclet number $\Pen_r$ in the high-P\'eclet-number regime. Third, the model reveals a redistribution of axial solute transport between the effective convective and dispersive fluxes, resulting in a reduction of the effective convective transport velocity in the dispersion model. Together, these findings provide new fluid-mechanical insight into non-classical dispersion at high P\'eclet numbers and establish an explicit reduced-order framework for its description and interpretation.

\section{Formulation of the transport problem}\label{sec:problem}

\subsection{Breakdown of the classical Taylor--Aris dispersion models at high P\'eclet numbers}
We first consider the convection-diffusion equation in cylindrical coordinates, as illustrated in \cref{fig:problem} (a), following \cite{Aris1960}:

\begin{equation}
    \frac{\partial c}{\partial t} + u(r)\frac{\partial c}{\partial x} = D_m \frac{\partial^2c}{\partial x^2} + D_m \frac{1}{r}\frac{\partial}{\partial r} \left(r\frac{\partial c}{\partial r}\right),
    \label{eq:conv_diff}
\end{equation}
where $x$ and $r$ denote the axial and radial coordinates, respectively, and $c$ denotes the contrast-agent concentration. Here, $D_m$ is the molecular diffusion coefficient and $u$ is the axial flow velocity. The infusion of the contrast agent is represented by a Dirichlet inlet boundary condition, yielding an initial-boundary-value problem.

The following dimensionless variables are then introduced:

\begin{equation}
    y=\frac{r}{R}, \xi=\frac{x}{L}, \tau=\frac{tU}{L}, w=\frac{u}{U}, C=\frac{c}{C_0},
    \label{eq:non_dim_var}
\end{equation}
where $R$ and $L$ denote the conduit radius and length, respectively, and $U$ is the characteristic flow velocity defined by

\begin{equation}
    U = f_u \int_0^{\frac{1}{f_u}} \avg{u}(t)\,dt,
    \label{eq:U}
\end{equation}
where $f_u$ is the fundamental frequency of the pulsatile flow and $1/f_u$ is its corresponding period, and $\avg{\cdot}$ denotes the
cross-sectional averaging operator defined by

\begin{equation}
    \avg{f}=2\int_0^1 f(y)y\,dy.
    \label{eq:average}
\end{equation}
Subsequently, \cref{eq:conv_diff} is dimensionalised as

\begin{equation}
	\frac{\partial C}{\partial \tau} + w(y)\frac{\partial C}{\partial \xi} = \frac{\epsilon}{\Pen_r} \frac{\partial^2C}{\partial \xi^2} + \frac{1}{\epsilon \Pen_r} \frac{1}{y}\frac{\partial}{\partial y} \left(y\frac{\partial C}{\partial y}\right).
	\label{eq:conv_diff_ndim}
\end{equation}
Following the classical Taylor--Aris framework, the concentration and velocity fields are decomposed into cross-sectional averages and fluctuations:

\begin{equation}
	C(\tau,\xi,y)=\avg{C}(\tau,\xi)+\tilde{C}(\tau,\xi,y), w(y)=\avg{w}+\tilde{w}(y).
	\label{eq:c_decom}
\end{equation}
Substituting \cref{eq:c_decom} into \cref{eq:conv_diff_ndim} and applying the cross-sectional averaging operator \cref{eq:average} to both sides of the equation yields \citep{Taylor1953}

\begin{equation}
	\frac{\partial \avg{C}}{\partial \tau}
	+ \avg{w}\frac{\partial \avg{C}}{\partial \xi}
	+ \frac{\partial \avg{\tilde{w}\tilde{C}}}{\partial \xi}
	= \frac{\epsilon}{\Pen_r} \frac{\partial^2\avg{C}}{\partial \xi^2},
	\label{eq:conv_diff_avg}
\end{equation}
which forms the starting point for the classical Taylor--Aris dispersion analysis.

\cref{eq:conv_diff_avg} highlights that the evolution of the averaged concentration $\avg{C}$ is not closed, due to the presence of the shear-induced dispersive flux term $\avg{\tilde{w}\tilde{C}}$. The classical Taylor--Aris theory achieves closure by assuming that the fluctuation $\tilde{C}$ remains asymptotically small compared to $\avg{C}$, allowing an explicit expression for $\avg{\tilde{w}\tilde{C}}$ to be derived \citep{Taylor1953, Aris1956, Aris1960}. This procedure leads to the well-known effective 1-D dispersion model:

\begin{equation}
	\frac{\partial \avg{c}}{\partial t} + \avg{u} \frac{\partial \avg{c}}{\partial x} = \left(D_m+\frac{\avg{u}^2 R^2}{48D_m}\right) \frac{\partial^2 \avg{c}}{\partial x^2},
	\label{eq:taylor_disp}
\end{equation}
To account for deviations from the parabolic velocity profile, a correction factor $\gamma_w$ can be introduced to characterize the modification of the dispersion process. This leads to the extended model \citep{Marbach2019}:

\begin{equation}
	\frac{\partial \avg{c}}{\partial t} + \avg{u} \frac{\partial \avg{c}}{\partial x} = \left(D_m+\frac{{U^2 R^2}}{48D_m} \gamma_w\right) \frac{\partial^2 \avg{c}}{\partial x^2}.
	\label{eq:achilles_disp}
\end{equation}
For pulsatile blood flow, the correction factor $\gamma_w$ was specifically investigated and quantitatively characterized by \citet{Wang2025}. Throughout the remainder of this paper, $\gamma_w$ is therefore evaluated according to the formulation established in that work.
 The corresponding dimensionless form of \cref{eq:achilles_disp} is

\begin{equation}
	\frac{\partial \avg{C}}{\partial \tau} + \avg{w} \frac{\partial \avg{C}}{\partial \xi} = \frac{\epsilon}{\Pen_r}\left(1+\frac{{\Pen_r^2}}{48} \gamma_w\right) \frac{\partial^2 \avg{C}}{\partial \xi^2}.
	\label{eq:achilles_disp_ndim}
\end{equation}

However, in arterial flows, where axial convection is much more rapid than transverse diffusion, the radial P\'eclet number

\begin{equation}
    \Pen_r = \frac{UR}{D_m}
    \label{eq:pe}
\end{equation}
is large, and the assumption of asymptotically small concentration fluctuations breaks down due to insufficient transverse diffusion, leading to the breakdown of the classical Taylor--Aris dispersion model. The progressive breakdown of Taylor--Aris dispersion with increasing $\Pen_r$ is illustrated in \cref{fig:problem} (b). At high $\Pen_r$, the excessively large effective axial diffusion predicted by \cref{eq:achilles_disp_ndim} causes the prescribed inlet concentration to propagate almost instantaneously to the outlet, resulting in nearly overlapping inlet and outlet concentration-time curves. As shown in \cref{fig:problem} (b), the Taylor--Aris predictions for $\Pen_r \ge \Om{5}$, corresponding to clinical contrast-agent transport, essentially overlap the inlet concentration.
In this regime, the dispersive flux $\avg{\tilde{w}\tilde{C}}$ cannot be closed using classical asymptotic approaches, motivating an alternative representation that does not rely on the assumption of asymptotically small fluctuations.

\subsection{Reformulation of 1-D dispersion model}

In general, the dispersive flux $\avg{\tilde{w}\tilde{C}}$ contains both local linear contributions associated with spatial variations of the averaged concentration and nonlinear or non-local effects, including, in particular, contributions induced by boundary and initial conditions. The latter contributions are denoted by $\mathcal{N}_{\tilde{w}}[\avg{C}]$. The local linear component can be formally represented in the spirit of Kramers--Moyal-type expansions, which gives:

\begin{equation}
	\avg{\tilde{w}\tilde{C}} = \theta_1\avg{C} + \theta_2\frac{\partial\avg{C}}{\partial\xi} + \theta_3\frac{\partial^2\avg{C}}{\partial\xi^2} + \dots + \mathcal{N}_{\tilde{w}}[\avg{C}].
	\label{eq:disp_flux}
\end{equation}
where the contributions contained in $\mathcal{N}_{\tilde{w}}[\avg{C}]$ are generally not expected to admit a closed-form representation.
Substituting \cref{eq:disp_flux} into \cref{eq:conv_diff_avg} yields

\begin{equation}
	\begin{split}
		\frac{\partial \avg{C}}{\partial \tau}
		+ (\avg{w}&+\theta_1) \frac{\partial \avg{C}}{\partial \xi}
		= (\frac{\epsilon}{\Pen_r}-\theta_2) \frac{\partial^2\avg{C}}{\partial \xi^2} \\
		&- \theta_3\frac{\partial^3\avg{C}}{\partial \xi^3}
		- \theta_4\frac{\partial^4\avg{C}}{\partial \xi^4} - \dots + O(\mathcal{N}_{\tilde{w}}).
	\end{split}
	\label{eq:achilles_disp_corr_full}
\end{equation}

In this study, we focus on a minimal effective closure that captures the dominant macroscopic longitudinal transport behaviour while retaining a tractable 1-D form. Within the framework of generalized Taylor dispersion theory \citep{Frankel1989}, the higher-order derivative terms in \cref{eq:achilles_disp_corr_full} represent successive corrections to the leading-order longitudinal transport dynamics, while the nonlocal contribution associated with $\mathcal{N}_{\tilde{w}}$ accounts for additional unresolved transport effects. For the reduced representation considered here, these contributions are not assumed to vanish identically; rather, their leading effects are incorporated into the effective convection and dispersive terms retained in the closure. This yields a minimal Taylor--Aris-like 1-D representation capable of describing the dominant longitudinal transport behaviour without explicitly resolving the higher-order and nonlocal contributions. The continued observation of Taylor-dispersion-like behaviour at high P\'eclet numbers in numerical and experimental studies \citep{Bijeljic2004, Tsakiroglou2005, Boschan2008, Gekle2017, Zhang2019} further supports the use of such a reduced representation. Accordingly, in the following analysis, the higher-order derivative terms and the nonlocal contributions associated with $\mathcal{N}_{\tilde{w}}$ are neglected, yielding a minimal Taylor--Aris-like 1-D dispersion model.

A further requirement is that \cref{eq:achilles_disp_corr_full} should recover the classical Taylor--Aris model in \cref{eq:achilles_disp_ndim} in the low-P\'eclet-number limit, ensuring consistency of the model across regimes. This consideration motivates a reparametrization of the coefficient structure in \cref{eq:achilles_disp_corr_full} as

\begin{equation}
	\left\{
	\begin{aligned}
		\theta_1 &= (\theta_u-1)\avg{w},\\
		\theta_2 &= -\theta_d\frac{\epsilon \Pen_r}{48} \gamma_w.
	\end{aligned}
	\right.
	\label{eq:theta_1_2}
\end{equation}
Substituting \cref{eq:theta_1_2} into \cref{eq:achilles_disp_corr_full}, and neglecting higher-order terms as well as contributions of $O(\mathcal{N}_{\tilde{w}})$, yields

\begin{equation}
	\frac{\partial \avg{C}}{\partial \tau} + \theta_u\avg{w} \frac{\partial \avg{C}}{\partial \xi} = \frac{\epsilon}{\Pen_r}\left(1+\theta_d\frac{{\Pen_r^2}}{48} \gamma_w\right) \frac{\partial^2 \avg{C}}{\partial \xi^2},
	\label{eq:patroclus_disp_ndim}
\end{equation}
where the dimensionless coefficients satisfy $\theta_u \to 1$ and $\theta_d \to 1$ as $\epsilon \Pen_r \to 0$, such that the proposed model recovers the classical Taylor--Aris dispersion model in dimensionless form, namely \cref{eq:achilles_disp_ndim}. Restoring all dimensionless variables in \cref{eq:patroclus_disp_ndim} yields a generalized 1-D dispersion model:

\begin{equation}
	\frac{\partial \avg{c}}{\partial t} + \theta_u\avg{u} \frac{\partial \avg{c}}{\partial x} = D_m\left(1+\theta_d\frac{{\Pen_r^2}}{48} \gamma_w\right) \frac{\partial^2 \avg{c}}{\partial x^2},
	\label{eq:patroclus_disp}
\end{equation}
with the remaining task being the determination of the unknown parameters $\theta_u$ and $\theta_d$.

\section{Identification of closure structure}
\label{sec:identification}

\begin{figure}
	\centering
	
	\begin{minipage}{\textwidth}
		\centering
		\includegraphics[width=\linewidth]{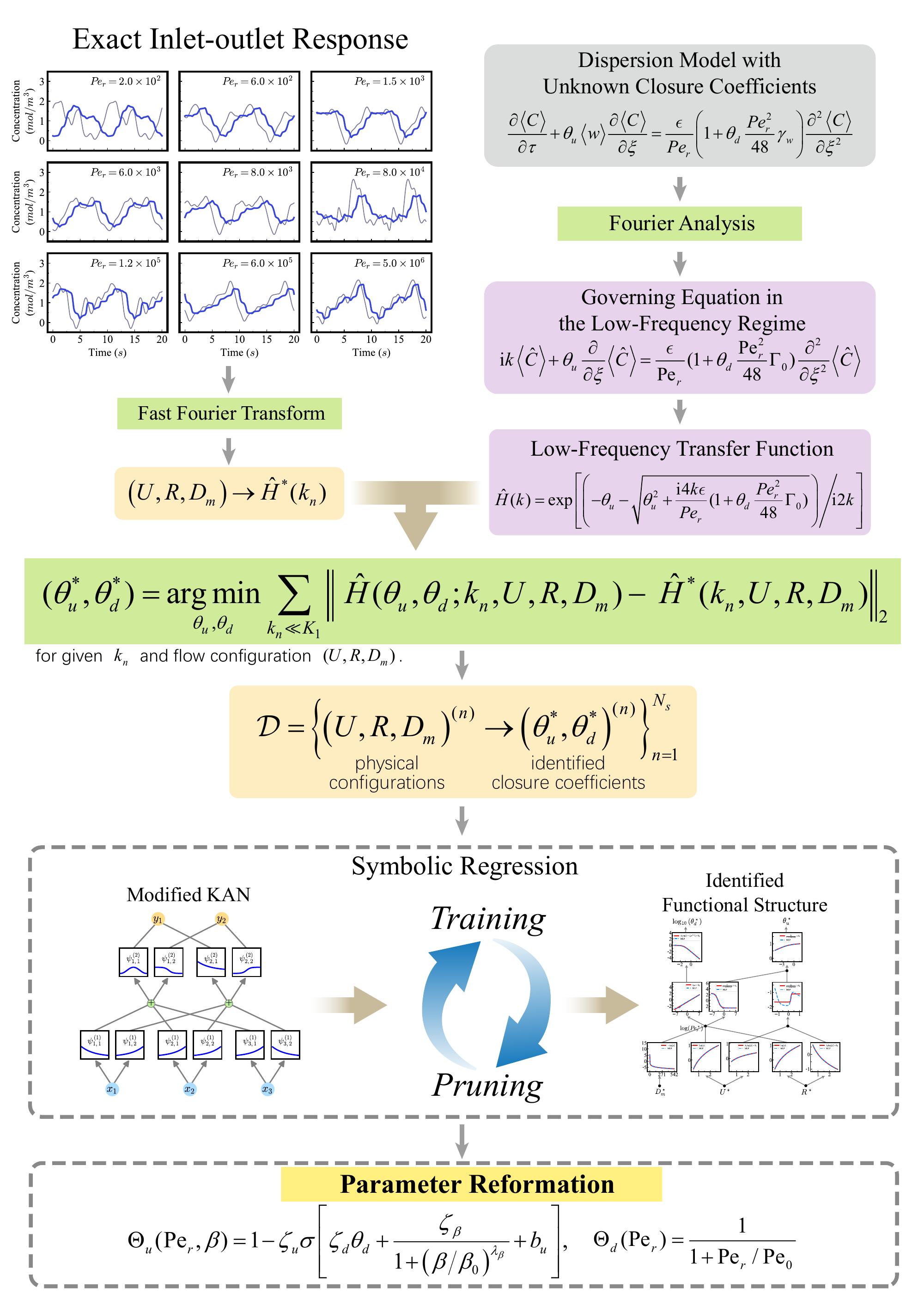}
	\end{minipage}
	
	\caption{Framework for identifying the functional structures of $\Theta_u$ and $\Theta_d$.}
	\label{fig:identification}
\end{figure}

The dimensional convection-diffusion \cref{eq:conv_diff} and the classical Taylor--Aris dispersion model \cref{eq:achilles_disp} involve three characteristic physical quantities, namely the molecular diffusivity $D_m$, the characteristic flow velocity $U$, and the conduit radius $R$. Accordingly, we assume that the closure coefficients depend on these characteristic quantities, such that

\begin{equation}
	\begin{cases}
		\theta_u = \Theta_u(D_m,U,R),\\
		\theta_d = \Theta_d(D_m,U,R).
	\end{cases}
	\label{eq:theta_u_d_map}
\end{equation}
The functional forms of $\Theta_u$ and $\Theta_d$, however, remain unknown. At high P\'eclet numbers, the breakdown of the classical Taylor--Aris asymptotic framework precludes the direct derivation of explicit closure relations using conventional analytical approaches. Nevertheless, obtaining an explicit representation of the closure structure is essential for the subsequent development and interpretation of the reduced-order transport model. This consideration motivates the adoption of an interpretable symbolic regression framework to identify the underlying closure structure $\Theta_u$ and $\Theta_d$.

This symbolic regression framework is illustrated in \cref{fig:identification}. The central idea is to transform the closure-identification problem into a coefficient-identification problem in the frequency domain. To this end, a Fourier-based analysis of \cref{eq:patroclus_disp_ndim} is first performed. By focusing on the low-frequency response, the contributions of higher-frequency components and their inter-frequency nonlinear interactions are neglected, leading to a quasi-linear transport system in which the input-output response can be represented by a transfer function. Subsequently, numerical solutions of the full convection-diffusion equation are obtained and analysed using the discrete Fourier transform (DFT), from which discrete samples of the corresponding transfer function are extracted, providing an exact representation of the inlet-outlet response of the full transport problem. The closure coefficients are identified by matching the transfer function predicted by the reduced quasi-linear system to that extracted from full numerical simulations. The resulting collection of identified closure coefficients then serves as the training dataset for a modified Kolmogorov--Arnold network (KAN), from which the underlying closure structure can be inferred. The methodology and results associated with each stage of the proposed identification framework are presented in the following subsections.

\subsection{Transfer function of the dispersion model in the low-frequency regime}
\label{sec:fourier_low_freq}

Since the subsequent identification procedure is performed in the frequency domain, Fourier representations of both $\avg{w}$ and $\gamma_w$ are first introduced. According to the non-dimensionalisation in \cref{eq:non_dim_var}, $\avg{w}$ can be expanded as a Fourier series with a mean value of 1:

\begin{equation}
	\avg{w}(t) = 1 + \sum_{n=1}^{\infty} W_n \cos(2\upi nf_u t + \phi^{(w)}_n),
	\label{eq:w_expan}
\end{equation}
where $W_n$ and $\phi^{(w)}_n$ denote the Fourier amplitudes and phases of $\avg{w}$ respectively and $f_u$ is the fundamental frequency. Using the dimensionless time defined in \cref{eq:non_dim_var}, \cref{eq:w_expan} becomes

\begin{equation}
	\avg{w}(\tau) = 1 + \sum_{n=1}^{\infty} W_n \cos(K_n \tau + \phi^{(w)}_n),
	\label{eq:w_expan_ndim}
\end{equation}
where $K_n=2\upi n\frac{f_uL}{U}$, denotes the $n$-th dimensionless angular frequency.
Following \citet{Wang2025}, $\gamma_w$ may be expressed as

\begin{equation}
	\gamma_w(t)=\left[1+\sum^{\infty}_{n=1}W_n\F{\gamma}_n(t; 1)\right] \left[1+\sum^{\infty}_{n=1}W_n\F{\gamma}_n(t; -1)\right].
	\label{eq:gamma_w}
\end{equation}
\begin{equation}
	\begin{split}
		\F{\gamma}_n(t; \sigma)&=\Gamma_A(\alpha_n)\e^{\iunit\sigma h_1(\alpha_n)t}\cos(2\upi nf_ut+\phi_n^{(w)})+ \\
		&\Gamma_B(\alpha_n)\e^{\iunit\sigma h_2(\alpha_n)t}\sin(2\upi nf_ut+\phi_n^{(w)}),
		\label{eq:gamma_wn}
	\end{split}
\end{equation}
where $\Gamma_A, \Gamma_B, h_1$ and $h_2$ depend only on the Womersley number $\alpha_n$ corresponding to the $n$-th frequency component of the pulsatile flow velocity. Since all frequency components of $\avg{w}$ are integer multiples of the fundamental frequency $K_1$, the nonlinear combinations appearing in $\gamma_w$ generate harmonics that remain integer multiples of $K_1$. Consequently, $\gamma_w$ is periodic with the same fundamental frequency and admits the Fourier representation:

\begin{equation}
	\gamma_w(\tau) = \Gamma_0 + \sum_{n=1}^{\infty} \Gamma_n \cos(K_n \tau + \phi^{(\gamma)}_n),
	\label{eq:gamma_expan_ndim}
\end{equation}
where $\Gamma_n$ and $\phi^{(\gamma)}_n$ denote the Fourier amplitudes and phases of $\gamma_w$ respectively.

\begin{figure}
	\centering
	
	\begin{subfigure}{\textwidth}
		\makebox[\linewidth][l]{(a)}
		\centering
		\includegraphics[width=\textwidth]{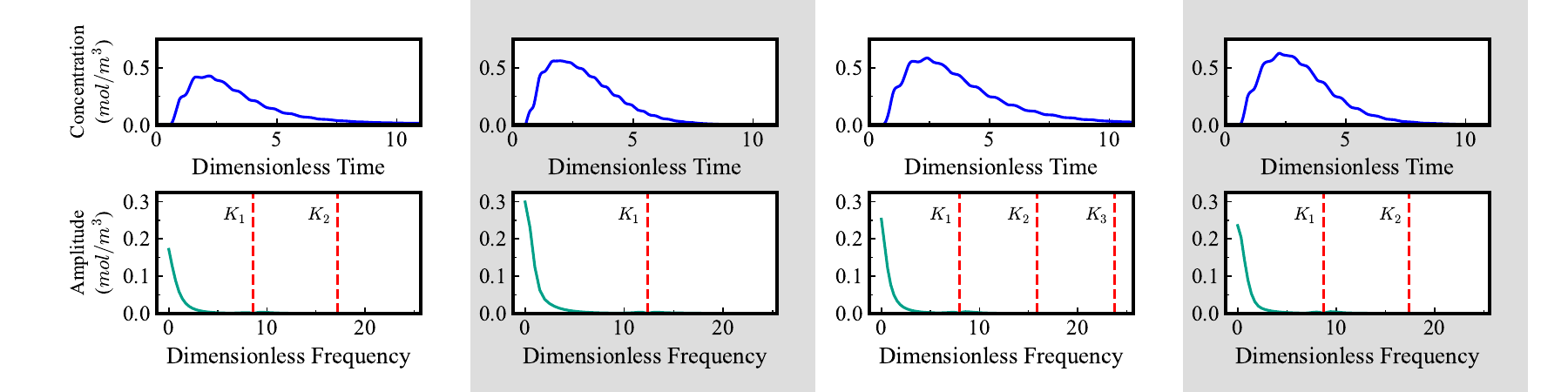}
	\end{subfigure}
	
	\vspace{0.5em}
	
	\begin{subfigure}{0.33\textwidth}
		\makebox[\linewidth][l]{(b)}
		\centering
		\includegraphics[width=\textwidth]{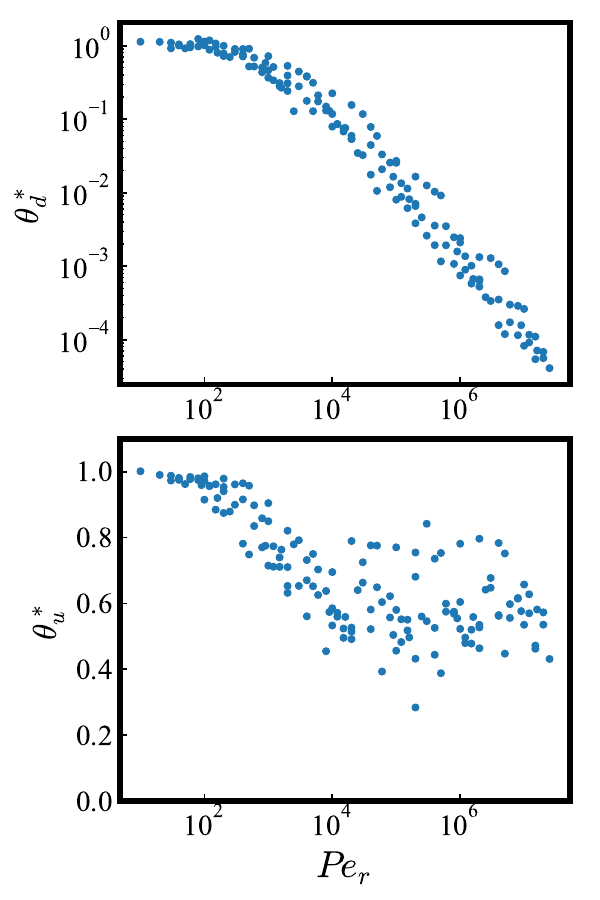}
	\end{subfigure}
	\hfill
	\begin{subfigure}{0.66\textwidth}
		\makebox[\linewidth][l]{(c)}
		\centering
		\includegraphics[width=\textwidth]{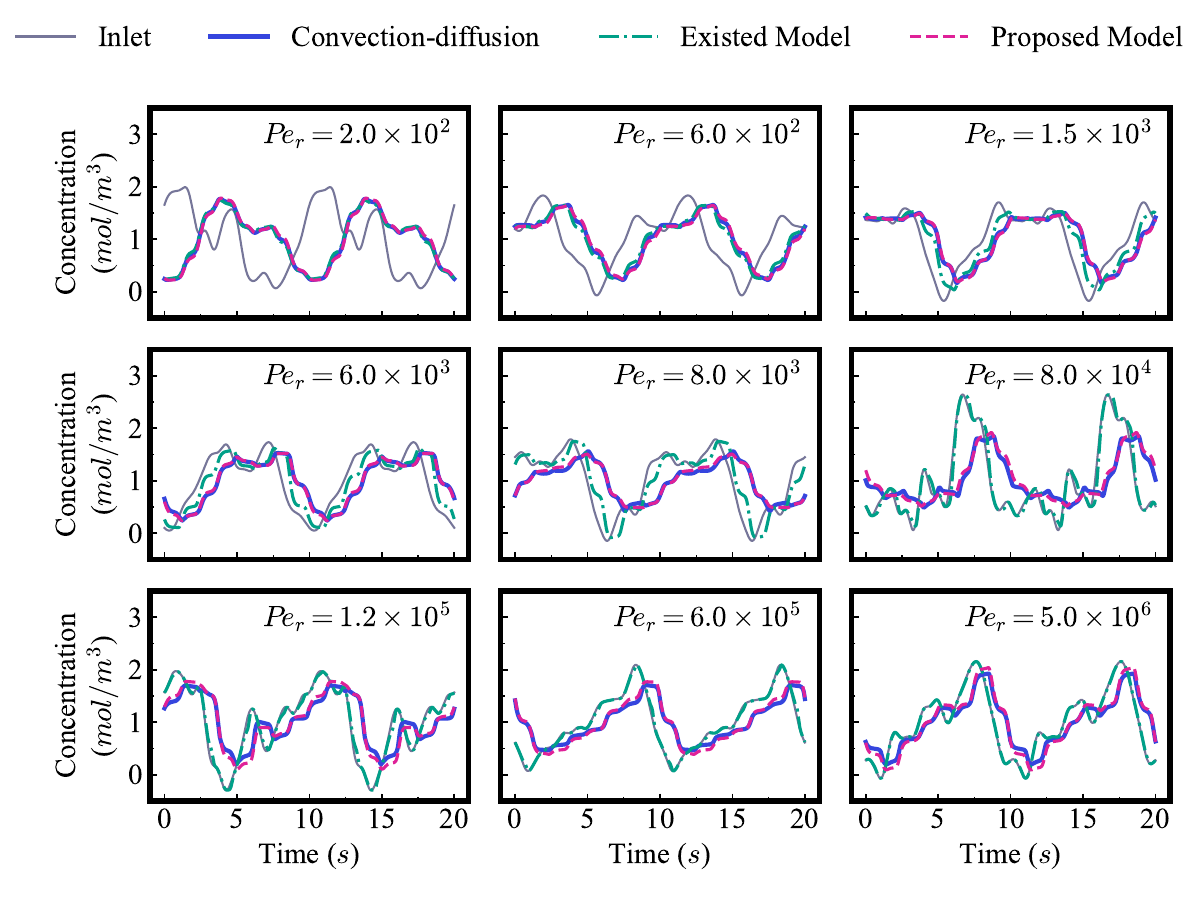}
	\end{subfigure}
	
	\caption{
		(a) Upper figures represent several cases of the numerical results for convection-diffusion equation with the concentration inlet boundary given by \cite{Ford2005}, which is approximated by a gamma-variate function. Lower figures show the DFT spectra of the corresponding numerical solutions. Most spectral energy is concentrated at frequencies much smaller than $K_1$, providing support for the low-frequency approximation employed in the derivation of \cref{eq:trans_func_steady}.
		(b) The identified values of $\theta_u^*$ and $\theta_d^*$ as functions of $\Pen_r$, revealing a strong dependence of $\theta_d^*$ on the radial P\'eclet number $\Pen_r$, whereas no comparable dependence is observed for $\theta_u^*$.
		(c) Numerical results at $\xi=0.5$ for convection-diffusion equation, \cref{eq:achilles_disp_ndim} and \cref{eq:patroclus_disp_ndim} at different radial P\'eclet number $\Pen_r$. Grey solid lines denote the inlet concentration profile. Blue solid lines, cyan dashed-dotted lines and pink dashed lines represent the numerical results for convection-diffusion equation, \cref{eq:achilles_disp_ndim} and \cref{eq:patroclus_disp_ndim}, respectively.
	}
	\label{fig:trans_func_fit}
\end{figure}

Substituting \cref{eq:w_expan_ndim,eq:gamma_expan_ndim} into \cref{eq:patroclus_disp_ndim} yields

\begin{equation}
	\begin{split}
		\frac{\partial \avg{C}}{\partial \tau} &+ \theta_u \left[ 1 + \sum_{n=1}^{\infty} W_n \cos(K_n \tau + \phi^{(w)}_n) \right] \frac{\partial \avg{C}}{\partial \xi} \\
		&= \frac{\epsilon}{\Pen_r} \left[1+\theta_d\frac{{\Pen_r^2}}{48}\Gamma_0 + \theta_d\frac{{\Pen_r^2}}{48}\sum_{n=1}^{\infty} \Gamma_n \cos(K_n \tau + \phi^{(\gamma)}_n) \right] \frac{\partial^2 \avg{C}}{\partial \xi^2}.
	\end{split}
	\label{eq:patroclus_expan}
\end{equation}
Taking the Fourier transform of \cref{eq:patroclus_expan} with respect to time, defined as

\begin{equation}
	\Fourier{\avg{C}}=\F{\avg{C}} = \int_{-\infty}^{\infty} \avg{C}(\xi, \tau) e^{-i k \tau} \, d\tau.
	\label{fourier}
\end{equation}
According to the standard properties of the Fourier transform,

\begin{equation}
	\Fourier{\frac{\partial \avg{C}}{\partial \tau}} = \iunit k \F{\avg{C}},
	\label{eq:fourier_prop1}
\end{equation}
\begin{equation}
	\F{\avg{C}}(\xi, k)=\overline{\F{\avg{C}}(\xi, -k)}.
	\label{eq:fourier_prop2}
\end{equation}
Furthermore, for the harmonic modulation term,

\begin{equation}
	\Fourier{\cos(K_n \tau + \phi^{(w)}_n)\avg{C}} = \F{\Phi}^-_n + \F{\Phi}^+_n,
	\label{eq:fourier_prop3}
\end{equation}
here, the overline represents the complex conjugate and $\delta$ is used as a shorthand for the Dirac delta function:

\begin{equation}
	\begin{cases}
		\F{\Phi}^-_n = \upi e^{\iunit\phi^{(w)}_n} \delta(k - K_n) \ast \F{\avg{C}}, \\
		\F{\Phi}^+_n = \upi e^{-\iunit\phi^{(w)}_n} \delta(k + K_n) \ast \F{\avg{C}}
	\end{cases}
	\label{eq:phi}
\end{equation}
for all $n>0$. By using property \cref{eq:fourier_prop2} and convolution property of Dirac function in \cref{eq:phi} yields

\begin{equation}
	\begin{cases}
		\F{\Phi}^-_n = \upi e^{\iunit\phi^{(w)}_n} \overline{\F{\avg{C}}(K_n-k)}, \\
		\F{\Phi}^+_n = \upi e^{-\iunit\phi^{(w)}_n} \F{\avg{C}}(K_n+k)
	\end{cases}
	\label{eq:phi_conv}
\end{equation}
Taking the Fourier transform of both sides of \cref{eq:patroclus_expan} yields

\begin{equation}
	\begin{split}
		\iunit k\F{\avg{C}} &+ \theta_u \frac{\partial}{\partial \xi} \left[\F{\avg{C}} + \sum_{n=1}^{\infty} W_n (\Phi^-_n+\Phi^+_n)\right] \\
		&= \frac{\epsilon}{\Pen_r} \frac{\partial^2}{\partial \xi^2} \left[(1+\theta_d\frac{{\Pen_r^2}}{48}\Gamma_0)\F{\avg{C}} + \theta_d\frac{{\Pen_r^2}}{48}\sum_{n=1}^{\infty} \Gamma_n (\F{\Phi}^-_n+\F{\Phi}^+_n)\right].
	\end{split}
	\label{eq:patroclus_freq}
\end{equation}

\cref{eq:phi_conv} indicates that the harmonic modulation of the pulsatile flow couples the Fourier mode at frequency $k$ to neighbouring modes shifted by $\pm K_n$. Nevertheless, two physical observations motivate a low-frequency approximation. First, longitudinal solute transport in conduits behaves as a low-pass filtering process, causing high-frequency concentration fluctuations to be strongly attenuated. Second, physiological contrast-agent injections are predominantly composed of low-frequency components, as illustrated in \cref{fig:trans_func_fit} (a). Consequently, for frequencies satisfying $k\ll K_1$, the shifted modes $\F{\avg{C}}(K_n\pm k)$ are expected to be much smaller than the low-frequency component $\F{\avg{C}}(k)$, implying that the contributions of $\Phi_n^-$ and $\Phi_n^+$ can be neglected in \cref{eq:patroclus_freq}. The effect of the dominant frequency of the concentration signal on the prediction accuracy of this approximation is further examined in \cref{sec:valid_c_bound}.

Subsequently, neglecting $\F{\Phi}^-_n$ and $\F{\Phi}^+_n$ in \cref{eq:patroclus_freq} yields

\begin{equation}
	\iunit k\F{\avg{C}} + \theta_u \frac{\partial}{\partial \xi}\F{\avg{C}} = \frac{\epsilon}{\Pen_r}(1+\theta_d\frac{{\Pen_r^2}}{48}\Gamma_0) \frac{\partial^2}{\partial \xi^2}\F{\avg{C}}.
	\label{eq:patroclus_freq_steady}
\end{equation}
As a result, the original dispersion model \cref{eq:patroclus_disp_ndim}, which contains pulsatile variations in $\avg{w}$ and $\gamma_w$ and their associated inter-frequency interactions, is reduced in the low-frequency limit to a quasi-linear system, \cref{eq:patroclus_freq_steady}. This reduction indicates that the transfer behaviour of the low-frequency components is primarily governed by the steady components of $\avg{w}$ and $\gamma_w$.
\cref{eq:patroclus_freq_steady} is an ordinary differential equation with respect to $\xi$, whose general solution is

\begin{equation}
	\begin{split}
		\F{\avg{C}}=\F{\avg{C}}_0&\exp \left( \frac{-\theta_u-\sqrt{\theta_u^2 + \frac{\iunit 4k \epsilon}{Pe_r}(1+\theta_d\frac{{Pe_r^2}}{48}\Gamma_0)}}{\iunit 2k} \xi \right) +\\
		&\F{\avg{C}}_1\exp \left( \frac{-\theta_u+\sqrt{\theta_u^2 + \frac{\iunit 4k \epsilon}{Pe_r}(1+\theta_d\frac{{Pe_r^2}}{48}\Gamma_0)}}{\iunit 2k} \xi \right).
	\end{split}
\end{equation}
Applying the boundary conditions

\begin{equation}
	\begin{cases}
		\F{\avg{C}}|_{\xi=0}=\F{\avg{C}}_{\mathrm{in}}(k), \\
		\F{\avg{C}}|_{\xi\rightarrow\infty}=0
	\end{cases}
\end{equation}
yields solution for \cref{eq:patroclus_freq_steady}:

\begin{equation}
	\F{\avg{C}}=\F{\avg{C}}_{\mathrm{in}}\exp \left( \frac{-\theta_u-\sqrt{\theta_u^2 + \frac{\iunit 4k \epsilon}{Pe_r}(1+\theta_d\frac{{Pe_r^2}}{48}\Gamma_0)}}{\iunit 2k} \xi \right),
	\label{eq:patroclus_freq_steady_solve}
\end{equation}
where the exponentially growing solution branch corresponding to $\F{\avg{C}}_1$ is discarded. Substituting the concentration frequency response at the outlet $\F{\avg{C}}_{\mathrm{out}}=\F{\avg{C}}|_{\xi=1}$ yields the low-frequency transfer function:

\begin{equation}
	\F{H}(k;\theta_u,\theta_d,U,R,D_m)=\frac{\F{\avg{C}}_{\mathrm{out}}}{\F{\avg{C}}_{\mathrm{in}}}=\exp \left( \frac{-\theta_u-\sqrt{\theta_u^2 + \frac{\iunit 4k \epsilon}{Pe_r}(1+\theta_d\frac{{Pe_r^2}}{48}\Gamma_0)}}{\iunit 2k} \right)
	\label{eq:trans_func_steady}
\end{equation}
for $k \ll K_1$. \cref{eq:trans_func_steady} indicates that the low-frequency inlet-outlet response of the proposed dispersion model \cref{eq:patroclus_disp_ndim} is governed by the closure coefficients $\theta_u$ and $\theta_d$.

\subsection{Identification of the closure coefficients using low-frequency transfer-function matching}

To identify the closure coefficients $\theta_u$ and $\theta_d$, the low-frequency transfer function derived in \cref{eq:trans_func_steady} is matched against the transfer response extracted from numerical simulations of the full convection-diffusion equation. The numerical scheme employed in this subsection discretizes the convection-diffusion equation using finite differences on a structured spatio-temporal grid represented as $(\tau_h,\xi_i,y_j)$, where $h$, $i$ and $j$ denote the indices of time steps, axial and radial discrete coordinates respectively. Time integration is performed using a weighted implicit $\theta$-scheme with $\theta=0.55$. The axial convective term is approximated by a first-order upwind difference, whereas the axial and radial diffusive terms were discretized using second-order central differences. Symmetry at the centerline and no-flux conditions at the wall were imposed through the corresponding finite-difference boundary stencils. The dimensionless discretization intervals are denoted by $\Delta\tau$, $\Delta\xi$ and $\Delta y$. All parameters involved in numerical methodology are shown in \cref{tab:num_params_trans_func}. Application of the above scheme yields a discrete concentration field $C_{i,j}^h$.

\begin{table}
	\begin{center}
		\def~{\hphantom{0}}
		\begin{tabular}{lccc}
			Parameter&	Description&	Unit&	Value \\[3pt]
			$L$&	Axial conduit length&	$\mathrm{mm}$&	$400.0$ \\
			$R$&	Conduit radius&	$\mathrm{mm}$&	$1.0-5.0$ \\
			$U$&	Characterize flow velocity&	$\mathrm{mm/s}$&	$100.0-500.0$ \\
			$D_m$&	Molecular diffusivity&	$\mathrm{m^2/s}$&	$10^{-10}-10^{-5}$ \\
			$\mu$&	Blood viscosity&	$\mathrm{mPa\cdot s}$&	$3.0$ \\
			$\rho$&	Mass density&	$\mathrm{kg/m^{3}}$&	$1060.0$ \\ 
			$f_u$&	Fundamental frequency of flow velocities&	$\mathrm{Hz}$&	$0.8-2.0$ \\ 
			$f_c$&	Fundamental frequency of concentration inlet boundaries&	$\mathrm{Hz}$&	$0.1$ \\ 
			$\Delta\tau$&	Discretization interval of dimensionless time&	-&	$0.01$ \\
			$\Delta\xi$&	Discretization interval of dimensionless axial length&	-&	$0.0025$ \\
			$\Delta y$&	Discretization interval of dimensionless radius&	-&	$0.005$ \\
		\end{tabular}
		\caption{Numerical simulation parameters for transfer-function matching.}
		\label{tab:num_params_trans_func}
	\end{center}
\end{table}

The cross-sectional-averaged concentration $\avg{C}_i^h$ is obtained from $C_{i,j}^h$ by numerical cross-sectional averaging and transformed into the frequency domain using DFT, defined as

\begin{equation}
	F^*[\avg{C}_i^h] = \F{\avg{C}}_i^n = \sum_{h=1}^{N_\tau} \avg{C}_i^h \Delta\tau \exp(\iunit k_n \tau_h),
	\label{eq:dft}
\end{equation}
where $n$ denotes the index of the frequency sample points corresponding to the $n$-th discrete frequency $k_n$ and $N_\tau$ represents total time step count. Furthermore, the inlet-outlet response of $\avg{C}^h_i$ can be expressed as

\begin{equation}
	\F{H}^*(k_n;U,R,D_m) = \frac{\F{\avg{C}}_{N_\xi}^n}{\F{\avg{C}}_1^n},
	\label{eq:trans_func_num}
\end{equation}
where $N_\xi$ represents total discrete axial coordinates. By matching \cref{eq:trans_func_steady} to the inlet-outlet responses given by \cref{eq:trans_func_num}, the coefficient-identification problem for $\theta_u$ and $\theta_d$ may be transformed into an optimization problem:

\begin{equation}
	(\theta_u^*,\theta_d^*) = \underset{\theta_u,\theta_d}{\arg\min} \sum_{k_n \ll K_1} \left\| \F{H}(\theta_u,\theta_d;k_n,U,R,D_m) - \F{H}^*(k_n,U,R,D_m) \right\|_2
	\label{eq:l2_trans_func}
\end{equation}
for given $k_n$ and flow configuration $(U,R,D_m)$.

To improve the robustness of the transfer-function matching procedure, a dedicated inlet concentration signal is employed. Since the closure coefficients are identified through the low-frequency optimization problem \cref{eq:l2_trans_func}, the accuracy with which low-frequency information is extracted from the numerical solution directly affects the subsequent coefficient identification. To maximize the amount of usable low-frequency information available for transfer-function matching, Fourier-based inlet concentration boundary

\begin{equation}
	C_{\mathrm{in}}=\avg{C}_{\mathrm{in}}=1+\sum_{n=1}^{N}A_n\cos(k_n\tau+\phi_n^{(c)})
	\label{eq:c_bound_fourier}
\end{equation}
is employed in the numerical simulation instead of physiological injection profiles. $k_n$ is set to $2\pi n \frac{f_cL}{U}$ where $f_c$ is the fundamental frequency of the concentration inlet. Furthermore, $f_c$ is set to a small value of $0.1Hz$ relative to the pulsatile flow velocity frequencies, ensuring a dense sampling of the low-frequency range relevant to the identification procedure. Subsequently, the observation window is chosen as one period of the fundamental concentration frequency, $1/f_c$, such that all prescribed harmonics coincide exactly with the DFT sampling frequencies. This choice minimizes spectral leakage and improves the accuracy of the transfer-function estimation.

To ensure that the synthetic inlet signal retains the dominant spectral characteristics of physiological contrast-agent injections, the Fourier amplitudes $A_n$ are prescribed as monotonically decaying functions of frequency. Consequently, most signal energy remains concentrated in the low-frequency range while a sufficient number of harmonics are preserved for transfer-function estimation. Specifically, the amplitudes are assigned according to

\begin{equation}
	A_n=\frac{1}{[1+(2\pi nf_c)^2]^{\zeta}},
	\label{eq:c_amplitude}
\end{equation}
where $\zeta\in(0.5,1)$ is randomly sampled. This construction generates a family of inlet signals whose spectra exhibit low-pass characteristics similar to those observed in physiological gamma-variate injection profile given by \cite{Ford2005}. The characteristic velocity $U$, conduit radius $R$ and molecular diffusivity $D_m$ are varied independently within physiologically relevant ranges, which is shown in \cref{tab:num_params_trans_func}. Repeating the transfer-function matching procedure over all sampled flow configurations yields a dataset

\begin{equation}
	\mathcal{D}=\left\{\left(U, R, D_m\right)^{(n)} \rightarrow \left(\theta_u^*, \theta_d^*\right)^{(n)}\right\}_{n=1}^{N_s},
	\label{eq:dataset_kan}
\end{equation}
where $N_s$ denotes the total number of sampled flow configurations. \cref{eq:dataset_kan} serves as the basis for the subsequent identification of the closure structure.

The results of the transfer-function matching procedure are presented in \cref{fig:trans_func_fit}. The identified values $\theta_u^*$ and $\theta_d^*$ obtained from transfer-function matching with respect to $\Pen_r$ is illustrated in \cref{fig:trans_func_fit} (b), revealing a strong dependence of $\theta_d^*$ on the radial P\'eclet number $\Pen_r$ whereas no comparable dependence is observed for $\theta_u^*$. Both $\theta_u^*$ and $\theta_d^*$ converge to $1$ as $\Pen_r$ decreases in \cref{fig:trans_func_fit} (b). Moreover, numerical results for convection-diffusion equation, \cref{eq:achilles_disp_ndim} and \cref{eq:patroclus_disp_ndim} at different $\Pen_r$ are illustrated in \cref{fig:trans_func_fit} (c) with blue solid lines, cyan dashed-dotted lines and pink dashed lines, respectively. Closure coefficients in \cref{eq:patroclus_disp_ndim} are taken as the identified values $\theta_u^*$ and $\theta_d^*$ obtained from transfer-function matching. It's observed that as $\Pen_r$ increases, results for \cref{eq:achilles_disp_ndim} deviate from convection-diffusion equation gradually, and eventually become nearly indistinguishable from the inlet signal, whereas results for \cref{eq:patroclus_disp_ndim} remain in good agreement with convection-diffusion equation.

\begin{figure}
	\centering
	
	\begin{subfigure}{0.98\textwidth}
		\makebox[\linewidth][l]{(a)}
		\centering
		\includegraphics[width=\textwidth]{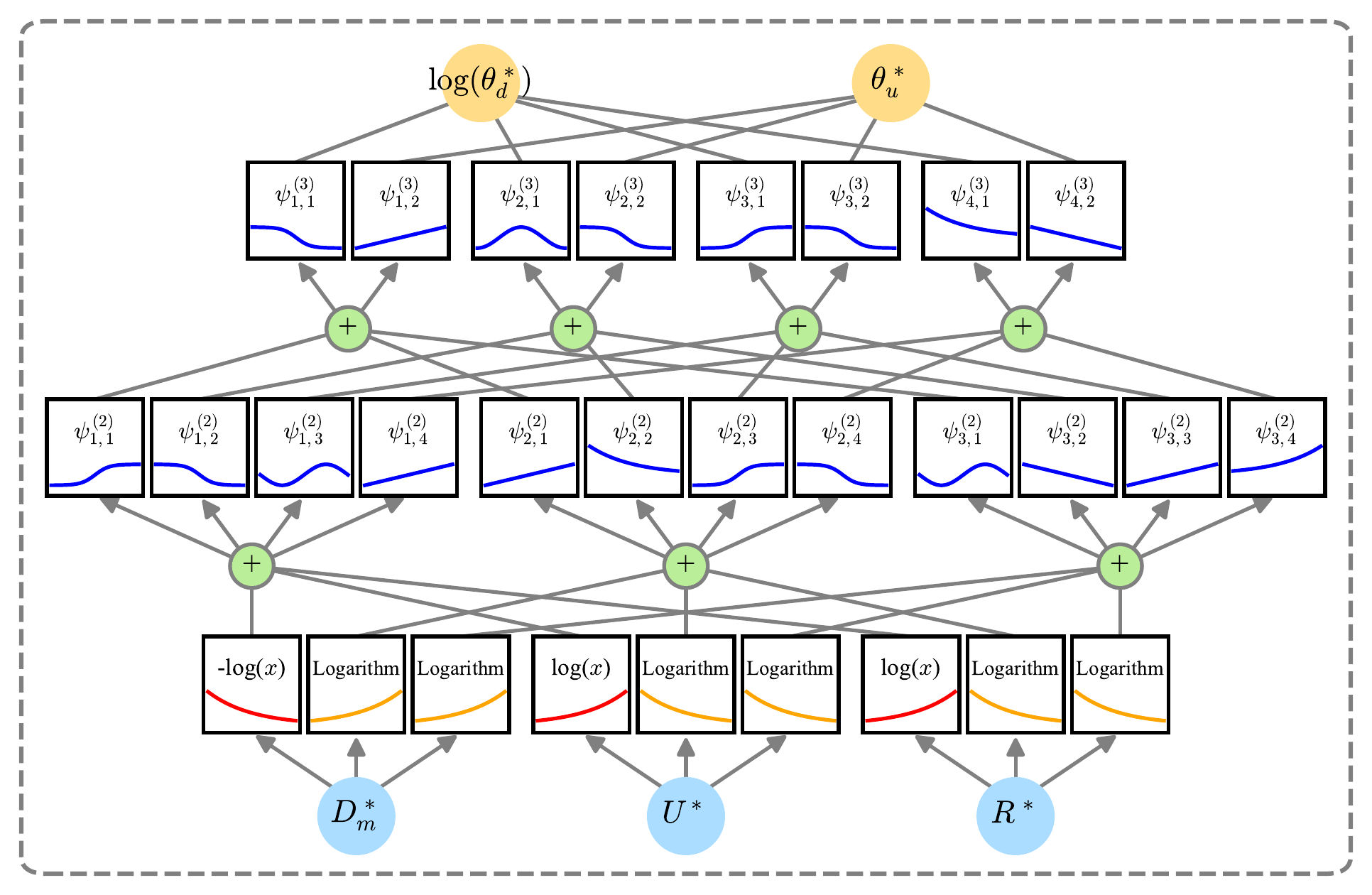}
	\end{subfigure}
	
	\vspace{0.5em}
	
	\begin{subfigure}{0.29\textwidth}
		\makebox[\linewidth][l]{(b)}
		\centering
		\includegraphics[width=\textwidth]{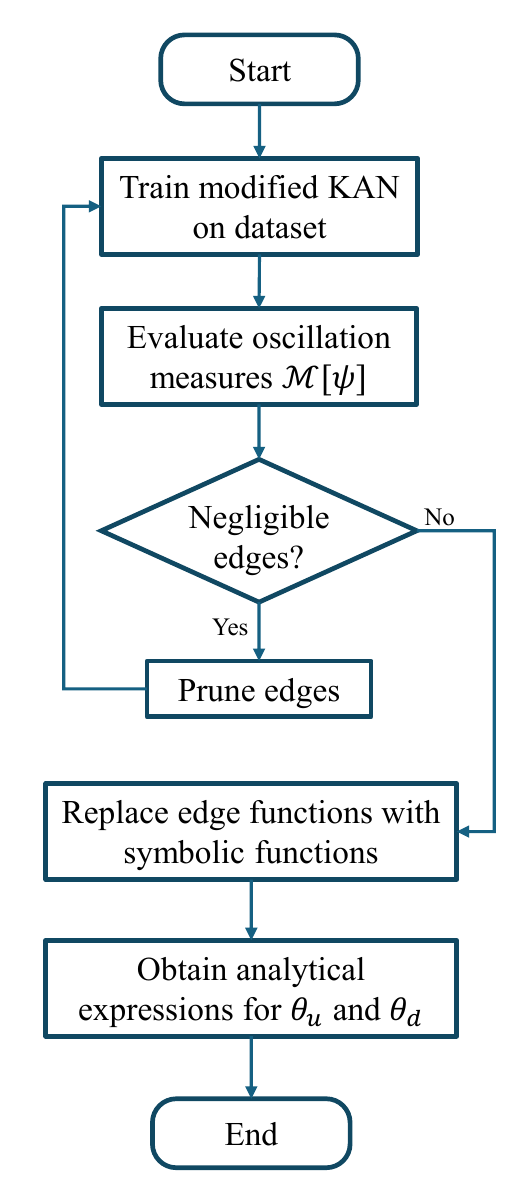}
	\end{subfigure}
	\hfill
	\begin{subfigure}{0.7\textwidth}
		\makebox[\linewidth][l]{(c)}
		\centering
		\includegraphics[width=\textwidth]{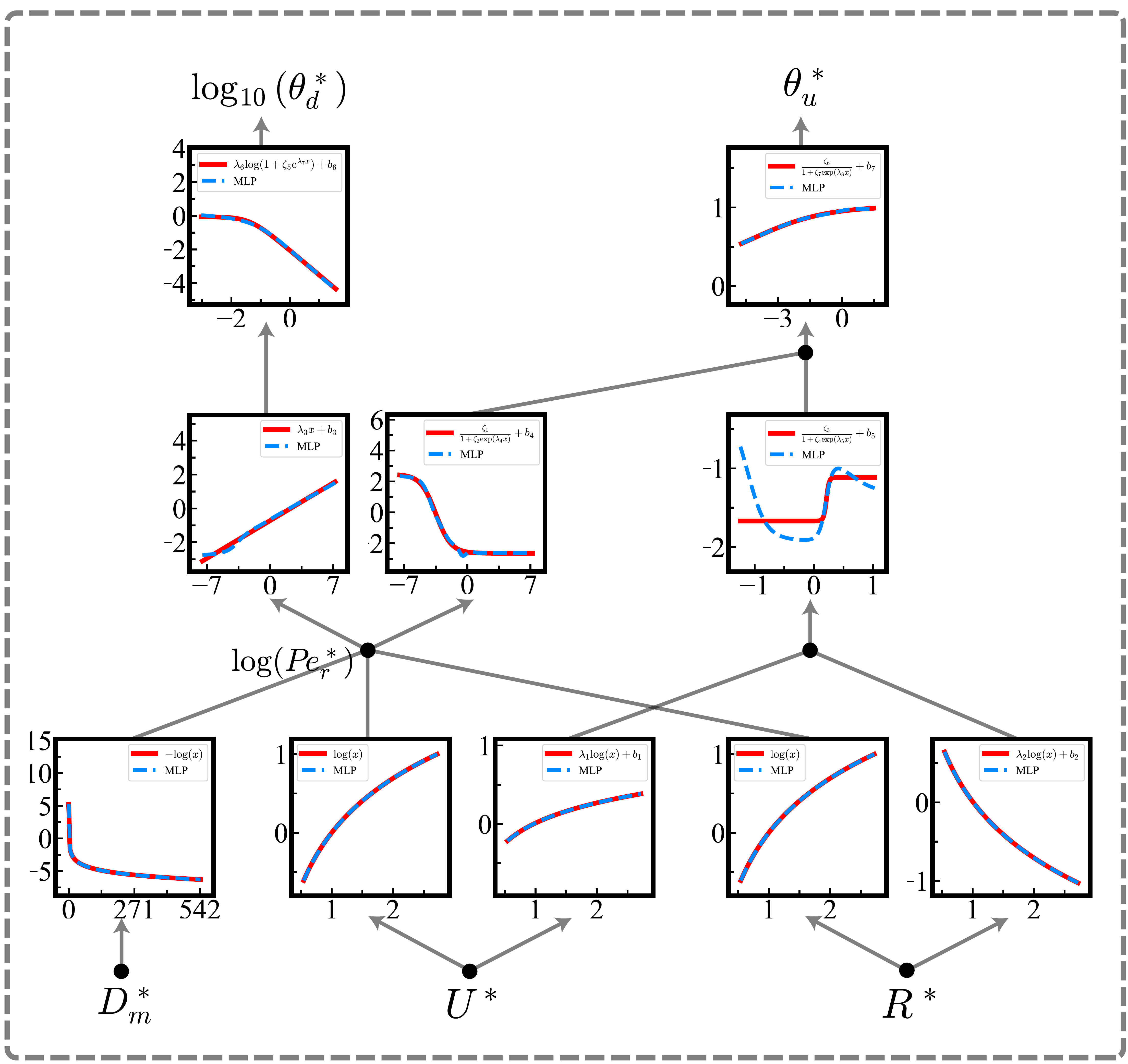}
	\end{subfigure}
	
	\caption{
		(a) Architecture of the modified KAN, including the MLP-based edge functions and logarithmic mappings introduced in the first hidden layer.
		(b) Workflow of symbolic structure discovery using modified KAN.
		(c) Final identified KAN structure after pruning and function replacement. Blue dashed lines denote the direct outputs of the MLP edge functions, whereas red solid lines represent their corresponding analytical replacements.
	}
	\label{fig:kan_plot}
\end{figure}

\subsection{Symbolic regression of the closure structure using Kolmogorov--Arnold networks}

Although the discrete mappings in \cref{eq:dataset_kan} provide numerical samples of $\theta_u$ and $\theta_d$, the underlying analytical forms of the mappings in \cref{eq:theta_u_d_map} remain unknown. To identify the underlying closure structure of $\Theta_u$ and $\Theta_d$, a KAN-based symbolic regression procedure is employed. KAN is an interpretable neural network architecture proposed by \cite{Liu2024}. Instead of relying entirely on a predefined symbolic library, KAN can adapt its functional building blocks directly to the data during training, allowing the functional structure to emerge without prescribing the specific functional forms a priori. At the same time, its neural-network formulation enables efficient gradient-based optimization and GPU acceleration, making the identification procedure computationally tractable while retaining an interpretable functional representation.
These features make KAN particularly attractive for closure identification, where the functional form is unknown a priori and may involve nonlinear interactions among multiple physical parameters.

Inspired by Kolmogorov--Arnold representation theorem, the forward propagation of KAN layers is defined as

\begin{equation}
	x^{(l+1)}_j=\sum_{i}^{N_I^{(l)}}\psi^{(l)}_{i,j}\left(x^{(l)}_i\right), j=1,\dots,N_J^{(l)},
	\label{eq:kan_layer}
\end{equation}
where $l$ denotes the index of a KAN layer, $N_{I}^{(l)}$ and $N_{J}^{(l)}$ denote input and output count of the $l$-th layer respectively. $\psi^{(l)}_{j,i}$ are edge functions with trainable parameters. All edge functions in \cref{eq:kan_layer} are one-to-one mappings which facilitates the analysis for functional structure. This property provides KAN with greater interpretability than most data-driven methods and, consequently, makes symbolic regression of $\Theta_u$ and $\Theta_d$ feasible.

This study adopts a modified KAN architecture, which is illustrated in \cref{fig:kan_plot} (a). The network consists of three KAN layers with a node configuration of $(3,3,4,2)$. The input variables are first normalized as

\begin{equation}
	D_m^*=\e \frac{D_m}{D_0}, U^*=\e \frac{U}{U_0}, R^*=\e \frac{R}{R_0}
	\label{eq:kan_input}
\end{equation}
corresponding to the characteristic quantities considered in \cref{eq:theta_u_d_map}, where $D_0$, $U_0$, and $R_0$ denote the statistical mean values of the corresponding variables over \cref{eq:dataset_kan}. The input and output dimensions are set to $3$ and $2$, respectively, corresponding to the three normalized input variables defined in \cref{eq:kan_input} and the two closure coefficients $\theta_u$ and $\theta_d$. In particular, the output corresponding to $\theta_d$ is represented in logarithmic form, since the values of $\theta_d^*$ span multiple orders of magnitude.

In the present study, three modifications are introduced to the KAN architecture compared to the original one. First, the spline-based edge functions are replaced by multi-layer perceptrons (MLPs) with 3 hidden layers of 48 nodes each, which provide stronger representation flexibility while improving the trainablity of the network. Second, the edge functions in the first hidden layer are initialized as trainable logarithmic mappings,

\begin{equation}
	\psi_{i,j}^{(1)}(x)=\lambda^{(1)}_{i,j}\log(x)+b^{(1)}_{i,j},
	\label{eq:log_edge}
\end{equation}
where $\lambda^{(1)}_{i,j}$ and $b^{(1)}_{i,j}$ are trainable parameters. The logarithmic form of \cref{eq:log_edge} introduces an inductive bias that encourages the network to construct multiplicative and divisive relationships among the input variables during symbolic regression. Finally, \cref{fig:trans_func_fit} (b) reveals a strong correlation between $\theta_d^*$ and the radial P\'eclet number $\Pen_r$. This observation suggests that explicitly introducing a functional representation of $\Pen_r$ into the network would facilitate the symbolic regression procedure. Therefore, the parameters of the edge functions at $l=1,j=1$ are fixed as $\zeta^{(1)}_{1,1}=-1$, $\zeta^{(1)}_{2,1},\zeta^{(1)}_{3,1}=1$ and $b^{(1)}_{1,1},b^{(1)}_{2,1},b^{(1)}_{3,1}=0$, which yields

\begin{equation}
	\begin{cases}
		\psi_{1,1}^{(1)}(x)=-\log(x), \\
		\psi_{2,1}^{(1)}(x)=\log(x), \\
		\psi_{3,1}^{(1)}(x)=\log(x).
	\end{cases}
\end{equation}
These fixed mappings explicitly encode the normalized radial P\'eclet number, namely

\begin{equation}
	\log(\Pen_r^*) = -\log(D^*_m)+\log(U^*)+\log(R^*),
	\label{eq:pe_encode}
\end{equation}
where

\begin{equation}
	\Pen_r^*=U^*R^*/D_m^*.
	\label{eq:pe_star}
\end{equation}
Consequently, functional dependencies on the radial P\'eclet number can be identified directly during symbolic regression, thereby improving both the efficiency of structure discovery and the physical interpretability of the resulting expressions.

The symbolic regression workflow adopted in the present study is described in \cref{fig:kan_plot} (b). Initially, all trainable parameters of the MLP edge functions are initialized using the Xavier initialization scheme. The network is then trained on the dataset by minimizing the $L_2$ loss function using the Adam optimizer. After convergence, a pruning procedure is performed to improve the compactness and physical interpretability of the resulting symbolic structure. To quantify the contribution of each edge function, the following oscillation measure is introduced:

\begin{equation}
	\mathcal{M}[\psi]=\sup_{x\in\Omega_x}\psi(x)-\inf_{x\in\Omega_x}\psi(x),
	\label{eq:edge_func_mag}
\end{equation}
where $\Omega_x$ denotes the domain of the corresponding input variable. \cref{eq:edge_func_mag} measures the variation of an edge function over its input domain. During the pruning procedure, all edge functions $\psi^{(l)}_{i,j}$ are examined, and their oscillation measures $\mathcal{M}[\psi^{(l)}_{i,j}]$ are compared with the other edge functions connected to the same output node $\left\{\psi^{(l)}_{i,j}\mid j=1,\dots,N^{(l)}_J\right\}$. If the oscillation measure of an edge function is significantly smaller than those of its neighboring edges, the corresponding edge is regarded as negligible and removed from the network by setting $\psi^{(l)}_{i,j}=0$. The pruned network is subsequently retrained on the dataset, and the pruning procedure is repeated until no further edge can be removed. At this stage, the symbolic structure is considered to have reached its maximum degree of compactness while preserving the predictive capability of the network. The remaining edge functions are subsequently replaced by analytical functions, yielding explicit expressions for the closure coefficients $\theta_u^*$ and $\theta_d^*$. The symbolic regression procedure described above is repeated multiple times, and the network yielding the lowest loss is selected as the final symbolic structure of the closure model. This procedure yields the following explicit expressions for the closure coefficients:

\begin{equation}
	\begin{cases}
		\theta_u^*=\frac{\zeta_6}{1+\zeta_7 \exp\left[
					\frac{\zeta_1\lambda_8}{1+\zeta_2{Pe_r^*}^{\lambda_4}} + \frac{\zeta_3\lambda_8}{1+\zeta_4 \exp(\lambda_5b_1+\lambda_5b_2) {U^*}^{\lambda_1\lambda_5}{R^*}^{\lambda_2\lambda_5}} + \lambda_8(b_4 + b_5)
			\right]} + b_7, \\
		\log_{10}(\theta_d^*)=\lambda_6\log(1+\zeta_5\e^{\lambda_7b_3}{\Pen_r^*}^{\lambda_3\lambda_7})+b_6,
	\end{cases}
	\label{eq:theta_u_d_kan}
\end{equation}
as illustrated in \cref{fig:kan_plot} (c).
\begin{table}
	\begin{center}
		\def~{\hphantom{0}}
		\begin{tabular}{cccc}
			Parameter&			Value&			Parameter&		Value	\\[3pt]
			$\lambda_1$&		~0.3714&		$b_4$&			-2.6502 \\
			$\lambda_2$&		-1.0364&		$b_5$&			-1.1151 \\
			$\lambda_3$&		~0.3172&		$b_6$&			-0.0649 \\
			$\lambda_4$&		~1.1792&		$b_7$&			~1.0196 \\
			$\lambda_5$&		44.6320&		$\zeta_1$&		~5.1096 \\
			$\lambda_6$&		-0.4621&		$\zeta_2$&		62.7122 \\
			$\lambda_7$&		~3.1528&		$\zeta_3$&		-0.5554 \\
			$\lambda_8$&		~0.5835&		$\zeta_4$&		7.64e-5 \\
			$b_1$&				~0.0116&		$\zeta_5$&		73.5568 \\
			$b_2$&				~0.0116&		$\zeta_6$&		-0.8360 \\
			$b_3$&				-0.7162&		$\zeta_7$&		11.7021 \\
		\end{tabular}
		\caption{Values of parameters in \cref{eq:theta_u_d_kan}.}
		\label{tab:struct_para_kan}
	\end{center}
\end{table}
The values of the parameters appearing in \cref{eq:theta_u_d_kan} are summarized in \cref{tab:struct_para_kan}.

Although \cref{eq:theta_u_d_kan} provides explicit analytical expressions for $\theta_u^*$ and $\theta_d^*$, the resulting parameterization remains unnecessarily complicated, hindering further mathematical analysis. We therefore perform an analytical simplification of the coefficient structure. The resulting compact expressions introduce the combined parameter

\begin{equation}
    \beta = \frac{R^3}{U},
    \label{eq:beta}
\end{equation}
and yield the following explicit closure relations:

\begin{equation}
\begin{cases}
        \theta_u=\Theta_u(\Pen_r, \beta) = 1 - \zeta_u \sigma\left[
            \zeta_d \theta_d + \frac{\zeta_{\beta}}{1+\left(\frac{\beta}{\beta_0}\right)^{\lambda_{\beta}}} + b_u
            \right], \\
        \theta_d=\Theta_d(\Pen_r) = \frac{1}{1+\frac{\Pen_r}{\Pen_0}}.
    \end{cases}
    \label{eq:theta_u_d_explicit}
\end{equation}
Details of the parameter reformulation are provided in Appendix A.
The parameters in the resulting expressions are further optimized against the original dataset. The optimized values are summarized in \cref{tab:struct_para_explicit}. Consequently, the unknown functional mappings postulated in \cref{eq:theta_u_d_map} are fully identified as the explicit analytical closure relations given in \cref{eq:theta_u_d_explicit}. These relations can now be directly incorporated into the proposed dispersion model \cref{eq:patroclus_disp_ndim}. The numerical behaviour and physical implications of the closure coefficients $\theta_u$ and $\theta_d$ are discussed in the following section, after which the predictive capability of the resulting closure model is validated.

\begin{table}
    \begin{center}
        \def~{\hphantom{0}}
        \begin{tabular}{lcc}
            Parameter&          Unit&                       Value\\
            $\Pen_0$&           -&                          1222.67\\
            $\beta_0$&          $\mathrm{m^2 s}$&           2.26e-8\\
            $\zeta_u$&          -&                          ~0.4699\\
            $\zeta_d$&          -&                          -4.6322\\
            $\zeta_{\beta}$&    -&                          -1.5459\\
            $\lambda_{\beta}$&  -&                          ~4.8019\\
            $b_u$&              -&                          ~2.1788\\
        \end{tabular}
    \end{center}
	\caption{Values of parameters in \cref{eq:theta_u_d_explicit}.}
	\label{tab:struct_para_explicit}
\end{table}

\section{Discussion and validation}

\subsection{Discussion on the proposed dispersion model}

\subsubsection{Functional characteristics of the closure coefficients}

\begin{figure}
	\centering
	
	\begin{subfigure}{0.325\textwidth}
		\makebox[\linewidth][l]{(a)}
		\centering
		\includegraphics[width=\textwidth]{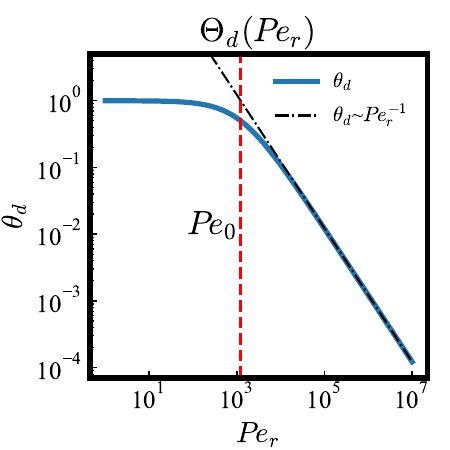}
	\end{subfigure}
    \begin{subfigure}{0.325\textwidth}
		\makebox[\linewidth][l]{(b)}
		\centering
		\includegraphics[width=\textwidth]{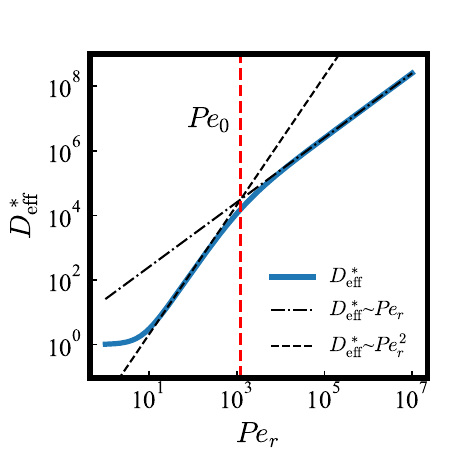}
	\end{subfigure}
    \begin{subfigure}{0.325\textwidth}
		\makebox[\linewidth][l]{(c)}
		\centering
		\includegraphics[width=\textwidth]{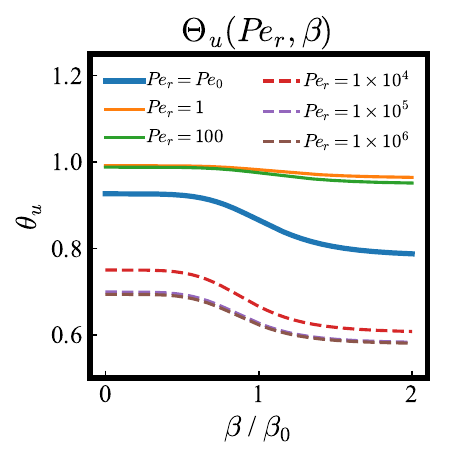}
	\end{subfigure}
	
	\vspace{0.5em}
	
	\begin{subfigure}{0.49\textwidth}
		\makebox[\linewidth][l]{(d)}
		\centering
		\includegraphics[width=\textwidth]{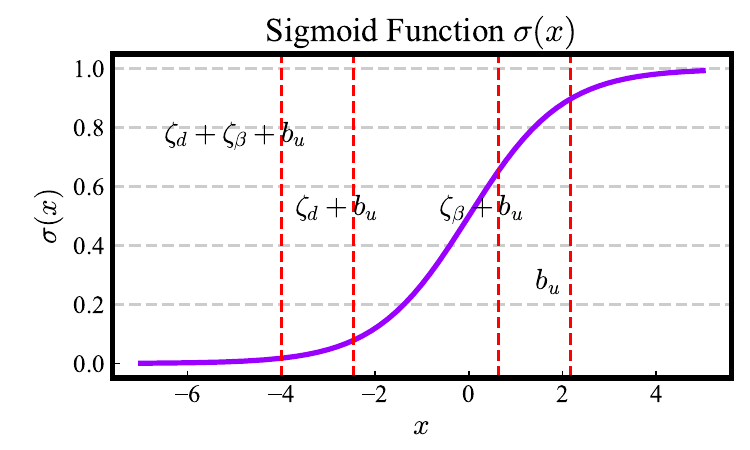}
	\end{subfigure}
	\hfill
	\begin{subfigure}{0.49\textwidth}
		\makebox[\linewidth][l]{(e)}
		\centering
		\includegraphics[width=\textwidth]{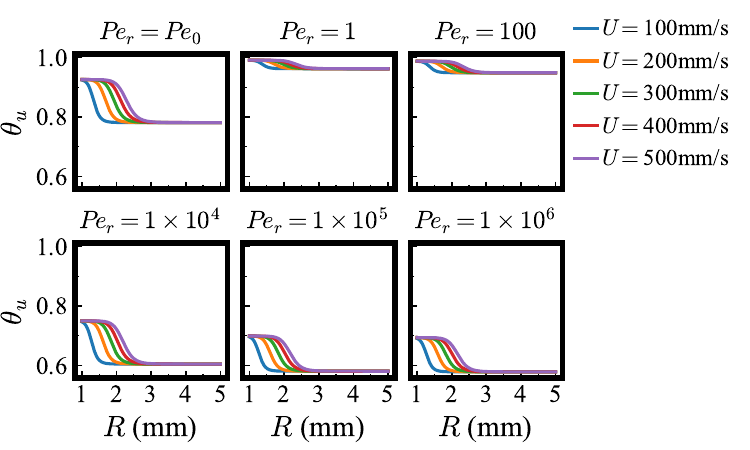}
	\end{subfigure}
	
	\caption{
		(a) The dependence of $\theta_d$ on the radial P\'eclet number $\Pen_r$. The red dashed line indicates the characteristic parameter $\Pen_0$, and the dash-dotted line represents the asymptotic scaling $\theta_d\sim\Pen_r^{-1}$.
        (b) The dependence of the dimensionless characteristic effective diffusion coefficient $D_{\mathrm{eff}}^*$ defined in \cref{eq:deff_patroclus_ndim} on $\Pen_r$. The red dashed line indicates the characteristic parameter $\Pen_0$. The dashed and dash-dotted lines represent the asymptotic scalings $D_{\mathrm{eff}}^*\sim\Pen_r^{2}$ and $D_{\mathrm{eff}}^*\sim\Pen_r$, respectively.
        (c) The dependence of $\theta_u$ on $\beta/\beta_0$ at $\Pen_r=\Pen_0$ and other representative radial P\'eclet numbers.
        (d) Illustration of the Sigmoid function $\sigma(\cdot)$. The four red dashed lines indicate the corresponding input values $\zeta_d+\zeta_{\beta}+b_u$, $\zeta_d+b_u$, $\zeta_{\beta}+b_u$, and $b_u$ of the Sigmoid function.
        (e) The dependence of $\theta_u$ on the conduit radius $R$ and characteristic flow velocity $U$ under different radial P\'eclet-number conditions.
	   }
	\label{fig:closure_coefficients}
\end{figure}
An analysis of the functional characteristics of the closure coefficients is performed in this subsection to provide further insight into their underlying behaviours and facilitate the subsequent physical interpretation. \cref{fig:closure_coefficients} (a) illustrates the dependence of $\theta_d$ on the radial P\'eclet number $\Pen_r$ according to \cref{eq:theta_u_d_explicit}. A pronounced variation of $\theta_d$ with $\Pen_r$ is observed, which is consistent with the strong correlation between the identified coefficient $\theta_d^*$ and $\Pen_r$ shown in \cref{fig:identification} (a). The characteristic parameter $\Pen_0$, indicated by the red dashed line in \cref{fig:closure_coefficients}(a), defines the transition scale governing the variation of $\theta_d$ with $\Pen_r$. In the low-P\'eclet-number regime, where $\Pen_r/\Pen_0\ll1$, $\theta_d$ approaches $1$, indicating a negligible correction to the classical dispersion coefficient. Conversely, in the high-P\'eclet-number regime, where $\Pen_r/\Pen_0\gg1$, the expression for $\theta_d$ asymptotically reduces to

\begin{equation}
    \theta_d = \frac{\Pen_0}{\Pen_r},
    \label{eq:theta_d_high_pe}
\end{equation}
demonstrating an inverse proportionality between $\theta_d$ and $\Pen_r$.

By taking $\gamma_w=1$ in \cref{eq:patroclus_disp}, the characteristic effective diffusion coefficient can be expressed as

\begin{equation}
    D_{\text{eff}} = D_m\left(1+\theta_d\frac{\Pen_r^2}{48}\right).
    \label{eq:deff_patroclus}
\end{equation}
The functional dependence of the dimensionless characteristic effective diffusion coefficient

\begin{equation}
    D_{\text{eff}}^* = \frac{D_{\text{eff}}}{D_m}
    \label{eq:deff_patroclus_ndim}
\end{equation}
on $\Pen_r$ is illustrated in \cref{fig:closure_coefficients} (b). The qualitative variation of $D_{\mathrm{eff}}^*$ with $\Pen_r$ agrees with the numerical and experimental results reported by \cite{Tsakiroglou2005} and \cite{Zhang2019}. In the low-P\'eclet-number regime, where $\Pen_r/\Pen_0\ll1$, $\theta_d$ approaches $1$ and \cref{eq:deff_patroclus} reduces to the classical Taylor--Aris dispersion relation:

\begin{equation}
    D_{\text{eff}} = D_m\left(1+\frac{\Pen_r^2}{48}\right),
    \label{eq:deff_taylor}
\end{equation}
which exhibits a second-order power-law scaling with $\Pen_r$. In contrast, in the high-P\'eclet-number regime, where $\Pen_r/\Pen_0\gg1$, substituting the asymptotic approximation given by \cref{eq:theta_d_high_pe} into \cref{eq:deff_patroclus} yields

\begin{equation}
    D_{\mathrm{eff}} = D_m\left(1+\frac{\Pen_r\Pen_0}{48}\right).
\end{equation}
When $\Pen_r\Pen_0/48\gg1$, the molecular diffusion contribution becomes negligible, and the effective diffusion coefficient further reduces to

\begin{equation}
    D_{\mathrm{eff}} = D_m\frac{\Pen_r\Pen_0}{48},
    \label{eq:deff_high_pe}
\end{equation}
demonstrating a transition from the classical quadratic scaling to a linear scaling of $D_{\mathrm{eff}}$ with $\Pen_r$.

The dependence of $\theta_u$ on the radial P\'eclet number $\Pen_r$ and the dimensionless variable $\beta/\beta_0$ is illustrated in \cref{fig:closure_coefficients} (c). The variation of $\theta_u$ with $\beta$ exhibits a switching behaviour. Specifically, $\theta_u$ remains approximately insensitive to $\beta$ when $\beta$ deviates substantially from the characteristic transition scale $\beta_0$, whereas a pronounced variation occurs in the vicinity of $\beta_0$. This behaviour originates from the Hill-type function embedded in the expression of $\theta_u$ in \cref{eq:theta_u_d_explicit}, where $\beta_0$ defines the transition threshold of the $\beta$-dependent contribution. The magnitude of this switching behaviour is further modulated by the radial P\'eclet number through $\theta_d$. In the low-P\'eclet-number regime, $\theta_u$ becomes relatively insensitive to $\beta/\beta_0$, whereas the strongest variation with $\beta/\beta_0$ occurs in the vicinity of $\Pen_r=\Pen_0$. $\theta_u$ approaches $1$ in the low-P\'eclet-number regime, indicating the recovery of the classical convective transport behaviour. Moreover, the overall magnitude of $\theta_u$ decreases as $\Pen_r$ increases.

The modulation behaviour of $\theta_u$ described above is governed by $\zeta_d$, $\zeta_{\beta}$ and $b_u$, as illustrated in \cref{fig:closure_coefficients} (d). In the low-P\'eclet-number regime, $\theta_d$ approaches 1, and the expression of $\theta_u$ in \cref{eq:theta_u_d_explicit} approximately reduces to

\begin{equation}
    \theta_u = 1 - \zeta_u \sigma\left[
        \frac{\zeta_{\beta}}{1+\left(\frac{\beta}{\beta_0}\right)^{\lambda_{\beta}}} + \zeta_d + b_u
    \right],
    \label{eq:theta_u_low_pe}
\end{equation}
which represents a Hill-type function of $\beta$ shifted by $\zeta_d+b_u$ and embedded within a Sigmoid function $\sigma(\cdot)$. The amplitude of this Hill-type contribution is $\zeta_{\beta}$, which determines the variation range of the Sigmoid input as $\beta$ changes. Specifically, the Sigmoid input varies within the interval $\left(\zeta_{\beta}+\zeta_d+b_u,\zeta_d+b_u\right)$. As illustrated in \cref{fig:closure_coefficients} (d), this interval is located in the saturated region of the Sigmoid function, resulting in a weak dependence of $\theta_u$ on $\beta$.
In contrast, in the high-P\'eclet-number regime, $\theta_d$ approaches 0, and \cref{eq:theta_u_d_explicit} approximately reduces to

\begin{equation}
    \theta_u = 1 - \zeta_u \sigma\left[
    \frac{\zeta_{\beta}}{1+\left(\frac{\beta}{\beta_0}\right)^{\lambda_{\beta}}} + b_u
    \right],
    \label{eq:theta_u_high_pe}
\end{equation}
where the Sigmoid input interval becomes $\left(\zeta_{\beta}+b_u,b_u\right)$. As illustrated in \cref{fig:closure_coefficients} (d), when the Sigmoid input enters the transition region bounded by $\zeta_d+b_u$ and $b_u$, the variation of $\theta_u$ with respect to $\beta$ becomes more pronounced.

\cref{fig:closure_coefficients} (e) illustrates the dependence of $\theta_u$ on the characteristic flow parameters $R$ and $U$ under different P\'eclet-number regimes. Similar to the dependence on $\beta$, $\theta_u$ exhibits a switching behaviour with respect to the conduit radius $R$. Specifically, $\theta_u$ remains relatively insensitive to $R$ when $R$ is far from the transition region, whereas $\theta_u$ decreases rapidly as $R$ increases across the transition region. The transition threshold of this switching behaviour is further modulated by the flow velocity $U$, with increasing $U$ shifting the transition towards larger values of $R$. In addition, the dependence of $\theta_u$ on $R$ and $U$ is strongly affected by $\Pen_r$. In both low- and high-P\'eclet-number regimes, $\theta_u$ exhibits weak sensitivity to the variation of $R$ and $U$, whereas pronounced dependence emerges in the vicinity of the transition scale $\Pen_0$. This behaviour is consistent with the modulation mechanism observed in the $\beta$-dependent functional behaviour of $\theta_u$.

\subsubsection{Flux redistribution mechanism of the proposed dispersion model}
\label{sec:discuss_interp}

\begin{figure}
	\centering
	
	\begin{subfigure}{\textwidth}
		\centering
        \makebox[\linewidth][l]{(a)}
		\includegraphics[width=\textwidth]{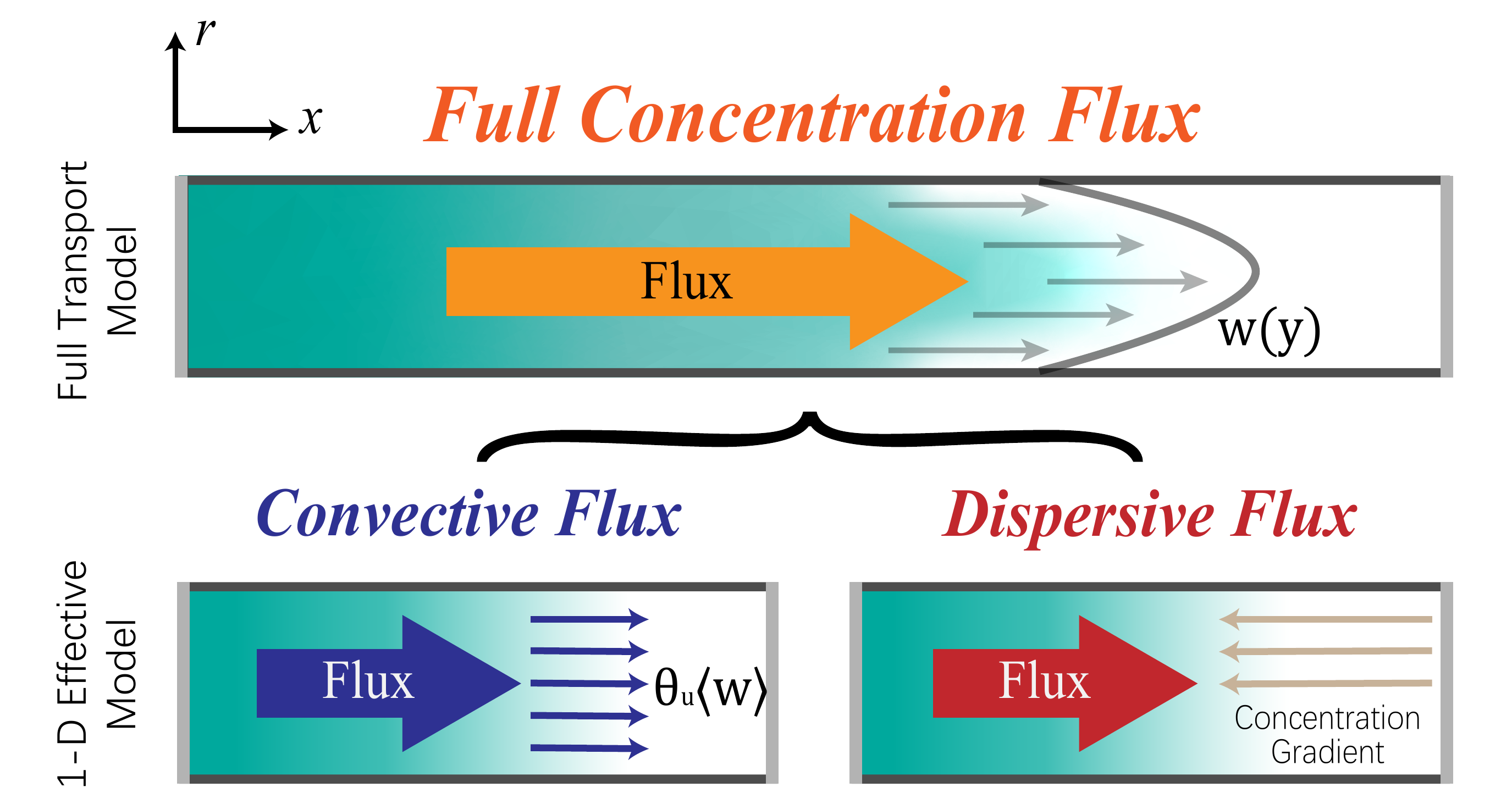}
	\end{subfigure}
	\hfill
    \begin{subfigure}{\textwidth}
		\centering
        \makebox[\linewidth][l]{(b)}
		\includegraphics[width=\textwidth]{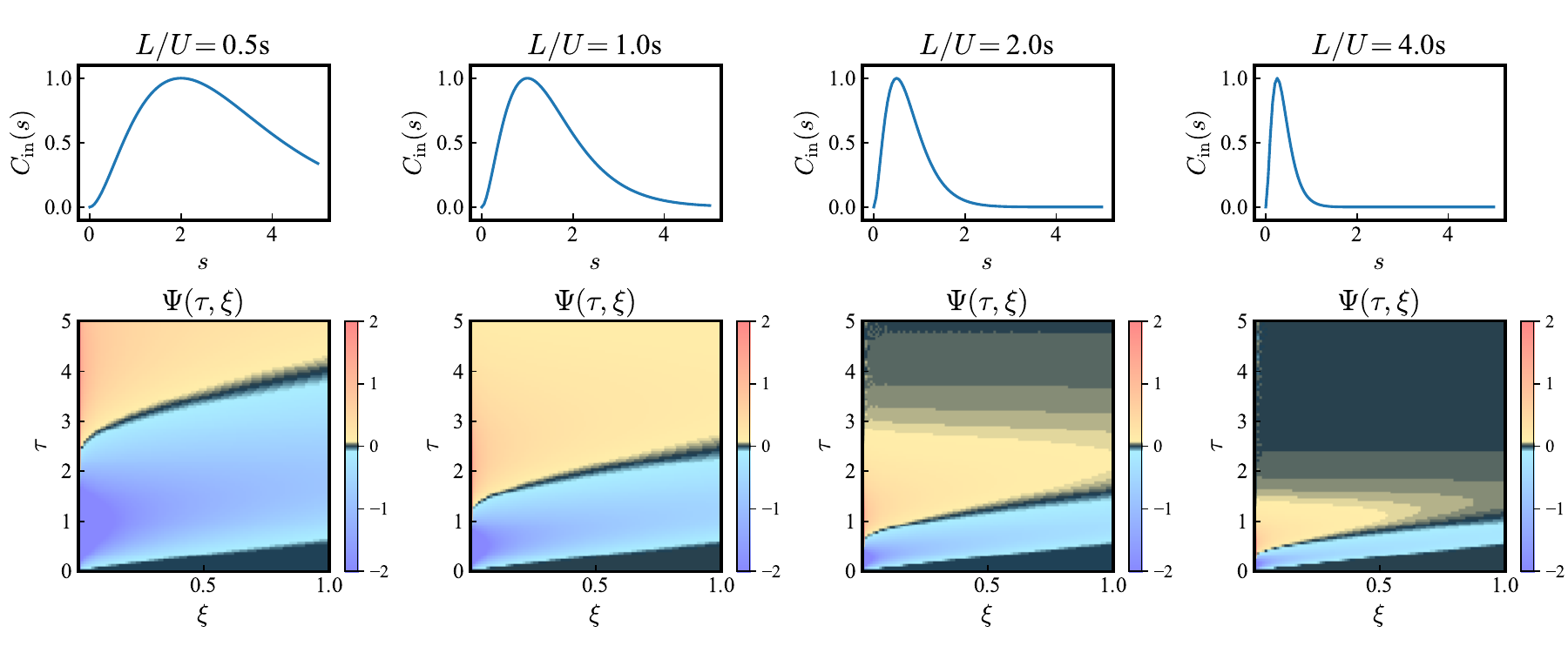}
	\end{subfigure}
	
	\caption{
        (a) The flux-redistribution mechanism underlying the proposed model. In the high-P\'eclet-number regime, the axial concentration flux in the full convection-diffusion model is dominated by convection. When this transport process is reduced to a 1-D description, the total convective concentration flux of the full model is redistributed between the effective convective and dispersive fluxes. The effective convective flux is driven by the modified mean velocity $\theta_u\avg{w}$, while the dispersive flux arises from axial diffusion driven by the mean concentration gradient $\frac{\partial\avg{C}}{\partial\xi}$, which is predominantly directed opposite to the downstream direction in the early-time regime.
		(b) Distribution of $\Psi(\tau,\xi)$ under different characteristic transport times $L/U$ for a prescribed pulse-like inlet concentration waveform. The upper panel illustrates the inlet concentration waveform, and the lower panels show the corresponding distributions of $\Psi(\tau,\xi)$ over the $(\tau,\xi)$. Dark- and light-blue regions indicate $\Psi(\tau,\xi)<0$, corresponding to the leading-edge region, whereas red and yellow regions indicate $\Psi(\tau,\xi)>0$, corresponding to the trailing-edge region. Regions where $\Psi(\tau,\xi)$ is close to zero are shown in black, separating the leading- and trailing-edge regions.
	}
	\label{fig:interpretation}
\end{figure}

The physical interpretation of the closure coefficients is discussed by considering the distinct dispersion regimes represented by low and high radial P\'eclet numbers. In the low-P\'eclet-number regime, the classical Taylor--Aris dispersion mechanism dominates. In terms of the closure coefficients, this regime is characterized by $\theta_u\rightarrow1$ and $\theta_d\rightarrow1$, under which the proposed dispersion model reduces to the classical Taylor--Aris formulation.

In contrast, at high radial P\'eclet numbers, radial diffusion becomes insufficient to fully homogenize the cross-sectional concentration distribution. The resulting dispersion behaviour becomes convection-dominated, where the contribution of molecular diffusion is negligible compared with the dispersion induced by the radially non-uniform velocity distribution.
According to \cref{eq:theta_u_d_explicit}, the transition between these two dispersion regimes is governed by $\Pen_0$. Since the classical Taylor--Aris dispersion theory has been extensively established in the low-P\'eclet-number regime, the following discussion focuses on the physical mechanisms underlying the convection-dominated dispersion behaviour, particularly the coupling between $\theta_u$ and $\theta_d$ in the regime where $\Pen_r>\Pen_0$.

Even when the classical Taylor--Aris dispersion becomes negligible, axial dispersion may still arise from the radial non-uniformity of the velocity field. The radial velocity gradient continuously stretches the concentration field in the streamwise direction, producing an axial broadening of the concentration distribution when the full transport process is represented by a 1-D effective model. This shear-induced stretching is therefore captured macroscopically by the dispersion term in the effective model. As $\Pen_r$ increases, the relative importance of advection over molecular diffusion increases, allowing the velocity shear to stretch the concentration field more strongly. Consequently, the resulting axial broadening becomes more pronounced, giving rise to an increasing dispersive contribution in the 1-D description.

Under the condition of convection-dominated dispersion, the axial cross-sectional concentration flux of the dimensionless convection-diffusion equation \cref{eq:conv_diff_ndim} is asymptotically dominated by the convective contribution, which can be expressed as

\begin{equation}
    J_w = \avg{wC}.
    \label{eq:c_flux_conv}
\end{equation}
Substituting the decomposition $w=\avg{w}+\tilde{w}$ and $C=\avg{C}+\tilde{C}$ into \cref{eq:c_flux_conv} gives

\begin{equation}
    J_w = \avg{\tilde{w}\tilde{C}} + \avg{w}\avg{C},
    \label{eq:c_flux_dicompose}
\end{equation}
where the cross terms $\avg{\tilde{w}\avg{C}}$ and $\avg{\avg{w}\tilde{C}}$ vanish due to the zero mean property of the perturbation terms. Substituting the dispersive flux expression \cref{eq:disp_flux} into \cref{eq:c_flux_dicompose}, while neglecting the nonlinear contribution $\mathcal{N}_{\tilde{w}}$ and higher-order derivative terms, gives

\begin{equation}
    J_w = (\theta_1 + \avg{w})\avg{C} + \theta_2\frac{\partial\avg{C}}{\partial\xi}.
    \label{eq:c_flux_theta_1_2}
\end{equation}
Subsequently, substituting the expressions of $\theta_1$ and $\theta_2$ given in \cref{eq:theta_1_2} into \cref{eq:c_flux_theta_1_2} yields

\begin{equation}
    J_w = \theta_u\avg{w}\avg{C} - \theta_d\frac{\epsilon\Pen_r\gamma_w}{48}\frac{\partial\avg{C}}{\partial\xi}.
    \label{eq:c_flux_theta_u_d}
\end{equation}
Furthermore, applying Fick's first law to \cref{eq:patroclus_disp_ndim} gives the diffusive concentration flux associated with the effective diffusion coefficient:

\begin{equation}
    J_\text{disp} = -\frac{\epsilon}{\Pen_r} \left(1+\theta_d\frac{\Pen_r^2}{48}\gamma_w\right) \frac{\partial\avg{C}}{\partial\xi}.
    \label{eq:fick_law}
\end{equation}
For convection-dominated dispersion, the axial molecular diffusion contribution becomes negligible, and \cref{eq:fick_law} reduces to

\begin{equation}
    J_{\text{disp}} = -\theta_d\frac{\epsilon\Pen_r\gamma_w}{48} \frac{\partial\avg{C}}{\partial\xi}.
    \label{eq:j_disp}
\end{equation}
Substituting \cref{eq:j_disp} into \cref{eq:c_flux_theta_u_d} yields

\begin{equation}
    J_w = \theta_u\avg{w}\avg{C} + J_\mathrm{disp}.
    \label{eq:c_flux_fick}
\end{equation}

\cref{eq:c_flux_fick} provides a conservative representation of the macroscopic concentration flux, establishing the relationship between the flux form of the convection-diffusion equation and the proposed dispersion model. The left-hand side is directly obtained from the dimensionless convection-diffusion equation \cref{eq:conv_diff_ndim}, whereas the right-hand side is derived from the asymptotic Kramers--Moyal-type expansion of the dispersive flux given by \cref{eq:disp_flux}.
For the convection-dominated dispersion described by the proposed model, the dispersive flux modified by $\theta_d$ and the convection flux modified by $\theta_u$ originate from the same cross-sectional convective transport flux $\avg{wC}$. Therefore, the convective and dispersive contributions in the proposed model are not independent corrections, but represent two competing components arising from the redistribution of the original convective transport during the 1-D equivalent simplification of the original transport process.

\cref{fig:interpretation} (a) illustrates the flux-redistribution mechanism underlying the proposed model. In the high-P\'eclet-number regime, the axial concentration flux in the full convection-diffusion model is dominated by convection. When this transport process is reduced to a 1-D description, the total convective concentration flux of the full model is redistributed between the effective convective and dispersive fluxes. The effective convective flux is driven by the modified mean velocity $\theta_u\avg{w}$, while the dispersive flux arises from axial diffusion driven by the mean concentration gradient $\frac{\partial\avg{C}}{\partial\xi}$.

In the early-time regime, the macroscopic concentration gradient is predominantly directed opposite to the downstream direction. Consequently, the resulting dispersive flux is directed downstream, in the same direction as the convective flux. The alignment of these two fluxes turns the redistribution into a competitive process: as dispersion becomes increasingly important with increasing $\Pen_r$, a progressively larger fraction of the total axial solute transport is represented by the dispersive flux, leaving a smaller fraction for the effective convective flux. As a result, the effective convective contribution is reduced, resulting in $\theta_u<1$, while the corresponding dispersive contribution is redistributed into $J_\text{disp}$.

Using \cref{eq:theta_u_d_explicit} in \cref{eq:j_disp} and performing straightforward algebraic rearrangement yields

\begin{equation}
    J_{\text{disp}} = \frac{\epsilon\Pen_0(\theta_d-1)\gamma_w}{48} \frac{\partial\avg{C}}{\partial\xi}.
    \label{eq:j_disp_reform}
\end{equation}
\cref{eq:j_disp_reform} reveals that the dependence of the dispersive flux on the radial P\'eclet number $\Pen_r$ is embedded through $\theta_d$. Since $\theta_d$ exhibits an approximately reciprocal dependence on $\Pen_r$, the magnitude of $J_{\mathrm{disp}}$ varies monotonically with $\Pen_r$ and gradually approaches an asymptotic limit.
The sign of the dependence between $J_{\mathrm{disp}}$ and $\Pen_r$ is determined by the concentration gradient $\frac{\partial\avg{C}}{\partial\xi}$. Although the concentration gradient may vary depending on the transient transport process and cannot be prescribed a priori, the functional behaviours of the closure coefficients provide an implicit constraint on its dominant trend. Specifically, the monotonic decrease of $\theta_u$ with increasing $\Pen_r$ revealed in \cref{fig:closure_coefficients} (c) and (e) indicates that the enhanced dispersive contribution must be accompanied by a reduction of the effective convective contribution. Therefore, the dominant concentration gradient associated with the convection-dominated dispersion described by the proposed model corresponds to $\frac{\partial\avg{C}}{\partial\xi}<0$, ensuring that the dispersive flux contributes in the downstream transport direction.

Consequently, as illustrated in \cref{fig:interpretation} (a), increasing $\Pen_r$ strengthens the competition between the dispersive flux $J_{\mathrm{disp}}$ and the effective convective contribution $\theta_u\avg{w}\avg{C}$, leading to a redistribution of the macroscopic concentration flux from the convective component towards the dispersive component. In the limit of $\Pen_r\gg\Pen_0$, substituting the asymptotic approximation given by \cref{eq:theta_d_high_pe} into \cref{eq:j_disp_reform} yields

\begin{equation}
    J_{\text{disp}} = -\frac{\epsilon\Pen_0\gamma_w}{48} \frac{\partial\avg{C}}{\partial\xi},
    \label{eq:j_disp_high_pe}
\end{equation}
indicating that the dispersive flux approaches an asymptotic high-P\'eclet-number limit and no longer increases with $\Pen_r$, which is consistent with the functional behaviours observed in \cref{fig:closure_coefficients} (c) and (e). Furthermore, the monotonic decrease of $\theta_u$ with increasing conduit radius $R$ observed in \cref{fig:closure_coefficients} (e) can be attributed to the geometric dependence of the dispersive flux through the aspect ratio $\epsilon$. Specifically, increasing $R$ increases $\epsilon$, which amplifies the dispersive contribution in \cref{eq:j_disp_reform,eq:j_disp_high_pe} and consequently reduces the fraction of the macroscopic flux represented by the effective convective component.

Moreover, the role of $\beta$ in \cref{eq:beta} can be understood from the range of characteristic parameters considered in this study. The molecular diffusivity $D_m$ spans a substantially wider range than the characteristic velocity $U$ and conduit radius $R$ encountered in clinical applications. Consequently, for a fixed high $\Pen_r$, variations in $U$ and $R$ are necessarily coupled: an increase in $U$ must be accompanied by a decrease in $R$, and vice versa. Their contributions to $\theta_u$ therefore act in opposite directions under a fixed-$\Pen_r$ constraint, indicating that the dependence of $\theta_u$ on $U$ and $R$ cannot be interpreted independently.

In this sense, $\beta$ provides a compact measure of the radius-dependent correction to the axial flux redistribution in the high-$\Pen_r$ regime. While $\Pen_r$ characterizes the overall relative importance of advection and molecular diffusion, $\beta$ captures how the redistribution of axial solute transport changes with the conduit radius at a given transport regime. The introduction of $\beta$ therefore allows the dependence of the effective convective contribution on the geometric scale of the conduit to be represented explicitly.

\subsubsection{Leading- and trailing-edge effects on convection-dominated dispersion}
\label{sec:lead_trail}

In physiological pulsatile blood flow, the inlet condition of contrast agents generally exhibits a pulse-like waveform, which can be described by the gamma-variate function introduced previously. Due to the characteristic temporal variation of the inlet concentration, the waveform can be divided into two distinct regions, namely the leading-edge and the trailing-edge. These two regions correspond to different concentration gradient evolutions during solute transport and may therefore induce distinct dispersive behaviours. In this section, the effects of the leading- and trailing-edges on the proposed dispersion model are investigated through analytical derivations, and the influence of the leading- and trailing-edge characteristics on the convection-dominated dispersion behaviour is elucidated.

In the convection-dominated regime, molecular diffusion is neglected, and the transport process is approximated by a purely convective equation. Furthermore, to isolate the macroscopic influence of the inlet concentration waveform, the effect of pulsatile velocity variation is neglected, and a steady parabolic velocity profile is considered.
Under these assumptions, the dimensionless convection-diffusion equation \cref{eq:conv_diff_ndim} reduces to

\begin{equation}
    \frac{\partial C}{\partial \tau} + 2(1-y^2)\frac{\partial C}{\partial \xi} = 0.
    \label{eq:conv_diff_ndim_nonconv}
\end{equation}
Imposing the Dirichlet boundary condition

\begin{equation}
    C(\tau,0,y)=C_{\mathrm{in}}(\tau)
    \label{eq:bound_analytical}
\end{equation}
at the inlet, the analytical solution of \cref{eq:conv_diff_ndim_nonconv} is obtained as

\begin{equation}
    C(\tau,\xi,y) = C_{\mathrm{in}} \left[ \tau-\frac{\xi}{2(1-y^2)} \right].
    \label{eq:c_analytical}
\end{equation}
Consequently, the cross-sectional averaged concentration is expressed as

\begin{equation}
    \avg{C}(\tau,\xi) = 2\int_0^1 y C_{\mathrm{in}} (s) dy,
    \label{eq:avg_c_analytical}
\end{equation}
where the characteristic variable $s$ is defined as

\begin{equation}
    s=\tau-\frac{\xi}{2(1-y^2)},
    \label{eq:chara_line}
\end{equation}
Indicating that the radial variation of the velocity profile $y$ can be mapped to different characteristic travel time $s$. Thus, the cross-sectional averaged concentration is determined by the superposition of inlet concentration inlet signal along $s$.

To facilitate the analysis of the leading- and trailing-edge effects, we define

\begin{equation}
    \Psi(\tau,\xi) = \frac{\partial\avg{C}}{\partial\xi},
    \label{eq:psi_c_grad}
\end{equation}
Substituting \cref{eq:avg_c_analytical} into \cref{eq:psi_c_grad} yields

\begin{equation}
    \Psi(\tau,\xi) = 2\int_0^1 y \frac{\partial C_{\mathrm{in}}(s)}{\partial s} \frac{\partial s}{\partial \xi} dy,
    \label{eq:psi_s}
\end{equation}
using \cref{eq:chara_line} in \cref{eq:psi_s} subsequently yields

\begin{equation}
    \Psi(\tau,\xi) = -\int_0^1 \frac{y}{1-y^2} g(y; \tau,\xi) dy,
    \label{eq:psi}
\end{equation}
where

\begin{equation}
    g(y; \tau,\xi) = \frac{\partial C_{\mathrm{in}}(s)}{\partial s}
    \label{eq:c_in_grad}
\end{equation}
denotes the local gradient of the inlet concentration signal with respect to the characteristic variable $s$, with $\tau$ and $\xi$ treated as parameters. \cref{eq:psi,eq:c_in_grad} indicate that, at a given spatio-temporal location $(\tau,\xi)$, the sign and magnitude of $\Psi$ are determined by the radial distribution of $g(y;\tau,\xi)$.

\cref{fig:interpretation} (b) illustrates the distribution of $\Psi(\tau,\xi)$ under different characteristic transport times $L/U$ for a prescribed pulse-like inlet concentration waveform. The upper panel shows the inlet concentration waveform, while the lower panels show the corresponding distributions of $\Psi(\tau,\xi)$ over the $(\tau,\xi)$ domain. Blue regions indicate $\Psi<0$, whereas red and yellow regions indicate $\Psi(\tau,\xi)>0$. Regions where $\Psi(\tau,\xi)$ is close to zero are shown in black, separating the two regions discussed above. At early transport times, the characteristic variable $s$ samples primarily the rising portion of the inlet waveform, where $\frac{\partial C_\mathrm{in}}{\partial s}>0$. Consequently, $g(y;\tau,\xi)$ is predominantly positive over the cross-section, and \cref{eq:psi} gives $\Psi<0$. As transport proceeds, the characteristic sampling progressively reaches the descending portion of the inlet waveform, where $\frac{\partial C_\mathrm{in}}{\partial s}<0$. The contribution of this region gradually increases and eventually dominates the cross-sectional integral, causing $\Psi(\tau,\xi)$ to transition from negative to positive. This transition indicates that the local concentration gradient evolves from being dominated by the rising edge of the inlet pulse to being dominated by its falling edge.

Analysis in \cref{sec:discuss_interp} reveals that the magnitude of the dispersive flux $J_{\mathrm{disp}}$ increases with increasing radial P\'eclet number $\Pen_r$ or conduit aspect ratio $\epsilon$. This dependence, together with the identified decrease in $\theta_u$, indicates that the dominant cross-sectionally averaged concentration gradient associated with the proposed convection-dominated dispersion is negative, namely $\Psi<0$. This motivates the definition of the leading-edge region as the spatio-temporal region $(\tau,\xi)$ satisfying $\Psi<0$, whereas the region satisfying $\Psi>0$ is defined as the trailing-edge region. The quantity $\Psi$ therefore provides a local measure of the dominant sign and magnitude of the concentration gradient under convection-dominated dispersion. In the leading-edge region, where $\Psi<0$, the dependence of $J_{\mathrm{disp}}$ on the governing transport parameters is qualitatively consistent with the dispersive behaviour identified by the proposed model. Accordingly, the leading-edge region represents transport conditions under which the flux-redistribution mechanism described by the proposed model is expected to be more pronounced.

Nevertheless, the local sign and magnitude of $\Psi$ may vary with the inlet concentration waveform, the spatio-temporal coordinates, and other transport conditions. The proposed model therefore describes an averaged dispersive behaviour, for which the applicable transport states are characterized by an overall downstream-directed dispersive tendency and consequently a predominantly negative $\Psi$. For the pulse-like contrast-agent transport considered in this study, such transport states are concentrated in the early-time portion of the concentration passage, where the leading edge dominates. As shown in \cref{fig:interpretation} (b), the fraction of the spatio-temporal domain occupied by the leading-edge region gradually decreases as the characteristic transport time $L/U$ increases, while the trailing-edge region becomes increasingly dominant. This reduction in the leading-edge-dominated regime suggests that the predictive accuracy of the proposed model may progressively deteriorate with increasing $L/U$, as the transport states become less representative of the early-time behaviour on which the model is primarily based. The dependence of model accuracy on the characteristic transport time $L/U$ is examined quantitatively in \cref{sec:valid_c_bound}.

\subsection{Validation of the proposed model}

\begin{table}
	\begin{center}
		\def~{\hphantom{0}}
		\begin{tabular}{lccc}
			Parameter&	Description&	Unit&	Value \\[3pt]
			$L$&	Axial conduit length&	$\mathrm{mm}$&	$200.0-400.0$ \\
			$R$&	Conduit radius&	$\mathrm{mm}$&	$1.0-5.0$ \\
			$U$&	Characterize flow velocity&	$\mathrm{mm/s}$&	$100.0-500.0$ \\
			$D_m$&	Molecular diffusivity&	$\mathrm{m^2/s}$&	$10^{-10}-10^{-6}$ \\
			$\mu$&	Blood viscosity&	$\mathrm{mPa\cdot s}$&	$3.0$ \\
			$\rho$&	Mass density&	$\mathrm{kg/m^{3}}$&	$1060.0$ \\ 
			$f_u$&	Frequency of the sinusoidal flow velocities&	$\mathrm{Hz}$&	$0.8-1.5$ \\ 
			$\Delta\tau$&	Discretization interval of dimensionless time&	-&	$0.01$ \\
			$\Delta\xi$&	Discretization interval of dimensionless axial length&	-&	$0.0025$ \\
			$\Delta y$&	Discretization interval of dimensionless radius&	-&	$0.005$ \\
		\end{tabular}
		\caption{Numerical simulation parameters for model validation.}
		\label{tab:num_params_valid}
	\end{center}
\end{table}

\begin{figure}
	\centering
	
	\begin{subfigure}{0.39\textwidth}
		\makebox[\linewidth][l]{(a)}
		\centering
		\includegraphics[width=\textwidth]{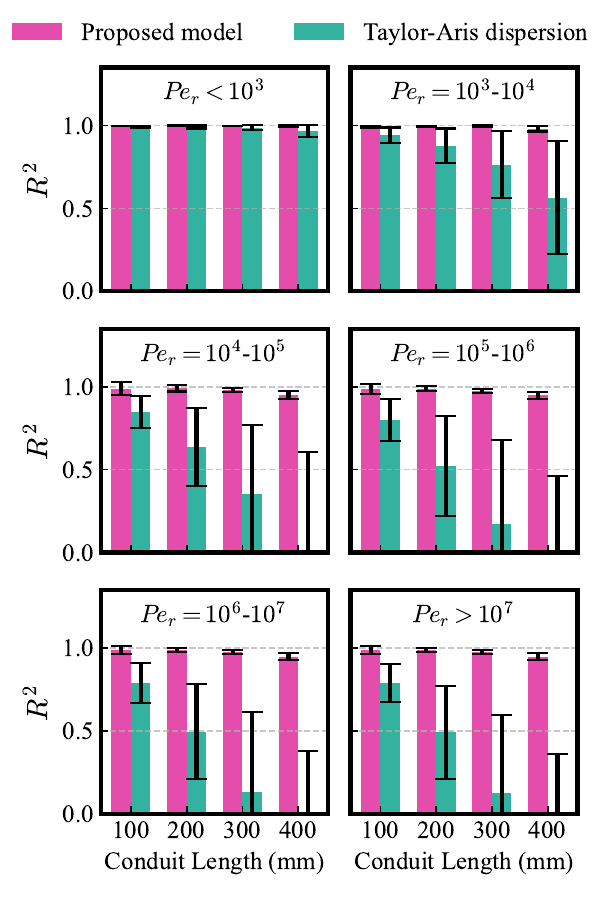}
	\end{subfigure}
	\hfill
    \begin{subfigure}{0.59\textwidth}
		\makebox[\linewidth][l]{(b)}
		\centering
		\includegraphics[width=\textwidth]{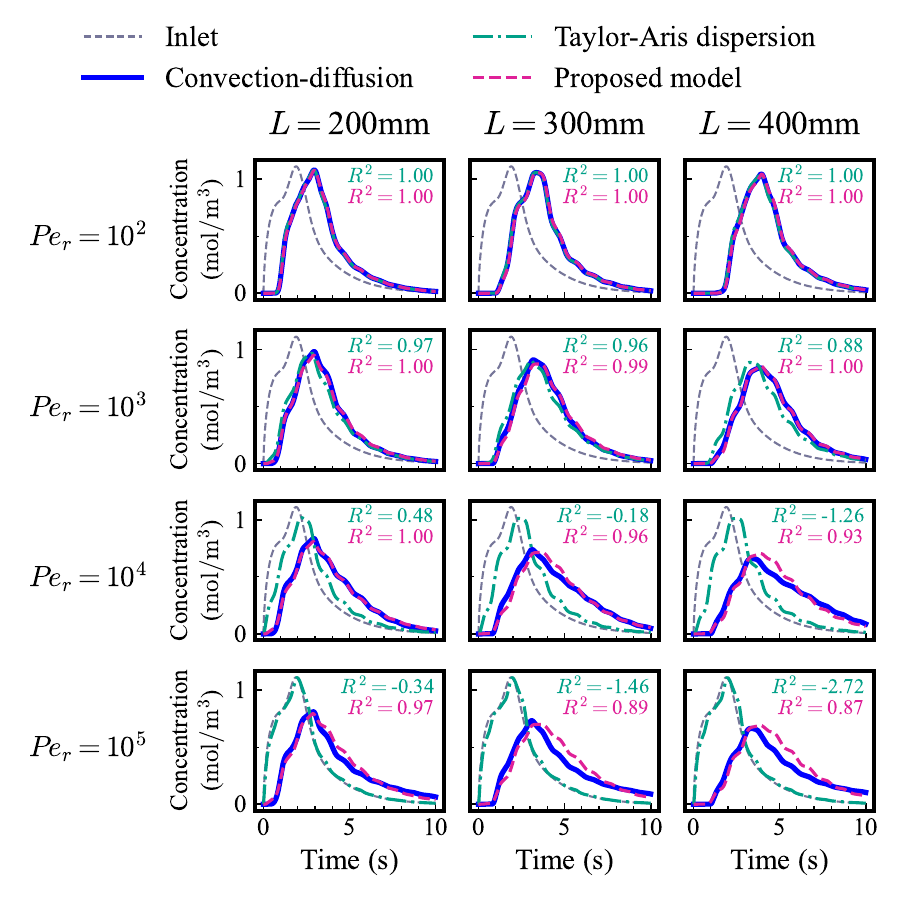}
	\end{subfigure}

	\caption{
		(a) The coefficients of determination obtained using \cref{eq:patroclus_disp} and \cref{eq:achilles_disp} under different configurations. The results obtained with \cref{eq:patroclus_disp} and \cref{eq:achilles_disp} are represented by pink and cyan bars, respectively, with the error bars indicating the standard deviation (STD). The results are further evaluated at different axial positions, with $L$ ranging from $100\mathrm{mm}$ to $400\mathrm{mm}$.
		(b) A representative set of numerical results under different radial P\'eclet numbers and axial positions. The twelve cases are arranged in four rows and three columns, with each row corresponding to a different radial P\'eclet number, $Pe_r=10^2$, $10^3$, $10^4$, and $10^5$, and each column corresponding to a different axial position, $L=200$, $300$, and $400,\mathrm{mm}$. The grey solid lines denote the inlet concentration profiles. The blue solid lines, cyan dash-dotted lines, and pink dashed lines represent the numerical solutions of the full convection-diffusion equation, \cref{eq:achilles_disp}, and \cref{eq:patroclus_disp}, respectively. The corresponding coefficients of determination, $R^2$, for the two effective models are also indicated in each case.
	}
	\label{fig:validation}
\end{figure}

\subsubsection{Dependence of the proposed model on the radial P\'eclet number}
\label{sec:valid_pe}

To validate the proposed dispersion model \cref{eq:patroclus_disp} over a broad range of radial P\'eclet numbers, numerical simulations are performed under different values of $Pe_r$. The numerical solutions of the full convection-diffusion equation, \cref{eq:conv_diff}, are taken as the ground truth. The proposed model \cref{eq:patroclus_disp} and the classical Taylor--Aris dispersion model \cref{eq:achilles_disp} are then solved numerically under the same conditions and compared with the ground-truth solutions. The coefficient of determination, $R^2$, is adopted as a quantitative measure of the predictive accuracy, defined as

\begin{equation}
    R^2=1-\frac{\sum(\avg{\widehat{C}}_i-\avg{C}_i)^2}{\sum(\avg{C}_i-\overline{\avg{C}})^2},
    \label{eq:deter_coef}
\end{equation}
where $\avg{C}_i$ denotes the cross-sectional averaged concentration at the $i$-th sampling point obtained from the ground-truth solution, $\overline{\avg{C}}$ represents the mean of all ground-truth values, and $\avg{\widehat{C}}_i$ denotes the corresponding cross-sectional averaged concentration predicted by the effective dispersion models.

Time integration is performed using a weighted implicit $\theta$-scheme with $\theta=0.55$. The axial convective term is approximated using a first-order upwind difference, whereas the axial and radial diffusive terms are discretized using second-order central differences. The symmetry condition at the centerline and the no-flux condition at the wall are imposed using the corresponding finite-difference boundary stencils. The physical parameters, including the conduit radius $R$, cross-sectional averaged flow velocity $U$, and molecular diffusion coefficient $D_m$, are randomly sampled within prescribed ranges for different numerical cases. All numerical parameters are summarized in \cref{tab:num_params_valid}, which broadly covers the range of physiological conditions considered in this study. 
The inlet concentration is prescribed using the gamma-variate function introduced previously:

\begin{equation}
    C_\mathrm{bound}(\tau)=C_0\tau^h e^{-h\tau},
    \label{eq:c_bound_valid}
\end{equation}
where $C_0$ and $h$ are randomly sampled within prescribed ranges for different numerical cases. The radial distribution of the axial velocity is prescribed according to the analytical Womersley solution for pulsatile flow \citep{Womersley1955}. The corresponding cross-sectionally averaged flow velocity is prescribed as a sinusoidal waveform:

\begin{equation}
	\avg{u}=U[1+0.8\sin(2\pi f_ut+\phi_u)].
	\label{eq:u_valid}
\end{equation}

The coefficients of determination obtained using \cref{eq:patroclus_disp} and \cref{eq:achilles_disp} under different configurations are shown in \cref{fig:validation} (a). The results obtained with \cref{eq:patroclus_disp} and \cref{eq:achilles_disp} are represented by pink and cyan bars, respectively, with the error bars indicating the standard deviation (STD). A total of $1800$ physical configurations are considered in \cref{fig:validation} (a), from which $7200$ cross-sectional concentration slices are sampled at four different axial positions, corresponding to $L=100\mathrm{mm}$, $200\mathrm{mm}$, $300\mathrm{mm}$, and $400\mathrm{mm}$, which further evaluates the results at different axial positions.
At low radial P\'eclet numbers, both models exhibit negligible discrepancies from the ground-truth solutions, resulting in consistently high coefficients of determination. As $Pe_r$ increases, however, the coefficient of determination of the Taylor--Aris model decreases markedly, whereas that of the proposed model remains relatively high. For both models, the coefficient of determination decreases as the axial position increases from $L=100\mathrm{mm}$ to $400\mathrm{mm}$. This degradation becomes more pronounced at high radial P\'eclet numbers. Notably, the proposed model exhibits a substantially smaller decrease in $R^2$ than the Taylor--Aris model.

A representative set of numerical results under different radial P\'eclet numbers and axial positions is presented in \cref{fig:validation} (b). The twelve cases are arranged in four rows and three columns, with each row corresponding to a different radial P\'eclet number, $Pe_r=10^2$, $10^3$, $10^4$, and $10^5$, and each column corresponding to a different axial position, $L=200$, $300$, and $400,\mathrm{mm}$. For each case, the inlet boundary condition and all other physical parameters are kept identical, such that the effects of $Pe_r$ and axial transport distance can be examined independently. The grey solid lines denote the inlet concentration profiles. The blue solid lines, cyan dash-dotted lines, and pink dashed lines represent the numerical solutions of the full convection-diffusion equation, \cref{eq:achilles_disp}, and \cref{eq:patroclus_disp}, respectively. The corresponding coefficients of determination, $R^2$, for the two effective models are also indicated in each case.

In \cref{fig:validation} (b), it's observed that at low radial P\'eclet numbers, both effective models closely overlap with the ground-truth solutions, with negligible differences between the two models. As $Pe_r$ increases, the prediction of the Taylor--Aris model increasingly deviates from the ground truth, whereas the proposed model remains in close agreement with the full convection-diffusion solution. With increasing axial transport distance, the agreement of both effective models with the ground truth gradually deteriorates. Nevertheless, the proposed model maintains a relatively high $R^2$ even at $L=400\mathrm{mm}$, whereas the degradation of the Taylor--Aris model becomes increasingly pronounced. Results above provide direct visual evidence that the proposed model retains high predictive accuracy over the range of physiological conditions considered, even at high radial P\'eclet numbers where the classical Taylor--Aris model exhibits substantial loss of accuracy. This behaviour is consistent with the analysis presented in the preceding sections and further demonstrates the robustness of the proposed model in the convection-dominated regime.

\subsubsection{Dependence of the proposed model on the inlet concentration signal}
\label{sec:valid_c_bound}

\begin{figure}
	\centering
	
	\begin{subfigure}{0.32\textwidth}
		\makebox[\linewidth][l]{(a)}
		\centering
		\includegraphics[width=\textwidth]{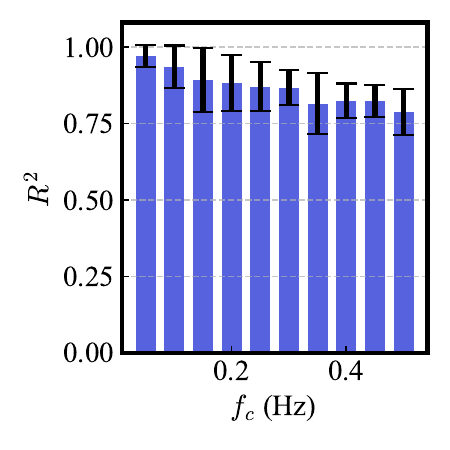}
	\end{subfigure}
	\hfill
    \begin{subfigure}{0.32\textwidth}
		\makebox[\linewidth][l]{(b)}
		\centering
		\includegraphics[width=\textwidth]{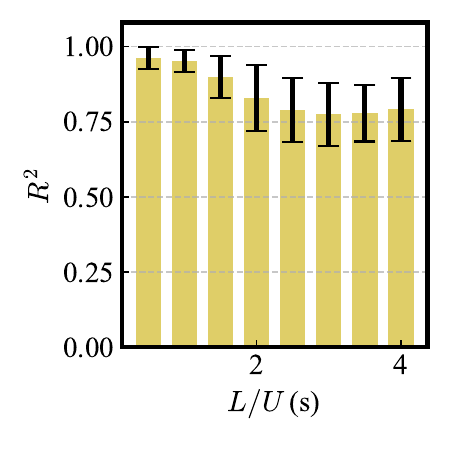}
	\end{subfigure}
    \hfill
    \begin{subfigure}{0.32\textwidth}
		\makebox[\linewidth][l]{(c)}
		\centering
		\includegraphics[width=\textwidth]{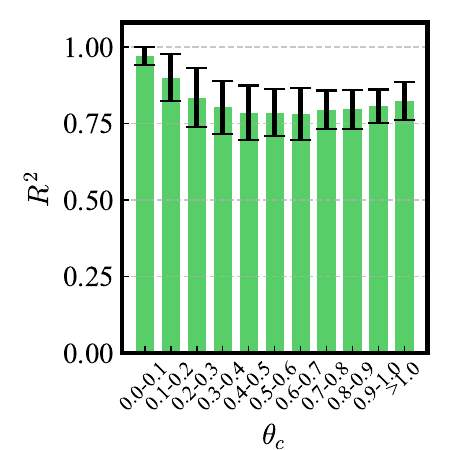}
	\end{subfigure}
	
	\vspace{0.5em}
	
	\begin{subfigure}{0.645\textwidth}
		\makebox[\linewidth][l]{(d)}
		\centering
		\includegraphics[width=\textwidth]{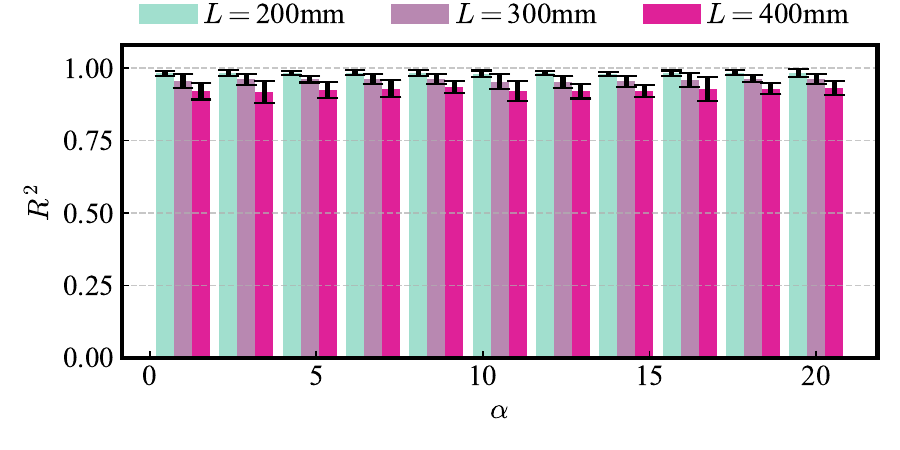}
	\end{subfigure}
	\begin{subfigure}{0.325\textwidth}
		\makebox[\linewidth][l]{(e)}
		\centering
		\includegraphics[width=\textwidth]{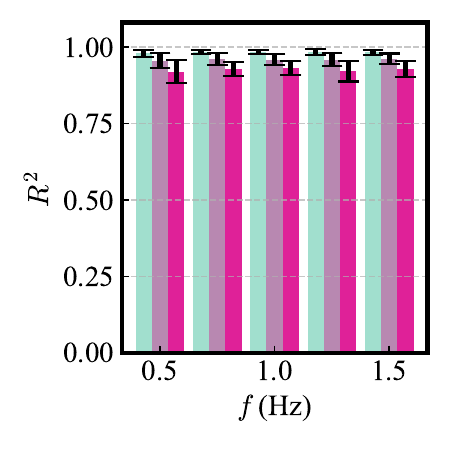}
	\end{subfigure}
	
	\caption{
		(a) Coefficients of determination between \cref{eq:patroclus_disp} and the ground-truth solutions under different fundamental inlet concentration frequencies, $f_c$.
        (b) Coefficients of determination between \cref{eq:patroclus_disp} and the ground-truth solutions under different characteristic transport times, $L/U$.
        (c) Coefficients of determination between \cref{eq:patroclus_disp} and the ground-truth solutions under different values of $\theta_c$, as defined by \cref{eq:theta_c}.
        (d) Coefficients of determination between \cref{eq:patroclus_disp} and the ground-truth solutions at three axial positions, $L=200\mathrm{mm}$, $300\mathrm{mm}$, and $400\mathrm{mm}$, as functions of the Womersley number, $\alpha$.
        (e) Coefficients of determination between \cref{eq:patroclus_disp} and the ground-truth solutions at three axial positions, $L=200\mathrm{mm}$, $300\mathrm{mm}$, and $400\mathrm{mm}$, as functions of the fundamental flow velocity frequency, $f_u$.
	}
	\label{fig:properties}
\end{figure}

\cref{sec:fourier_low_freq} derives a transfer function describing the low-frequency inlet-outlet concentration response, \cref{eq:trans_func_steady}, which is subsequently employed to identify the closure coefficients $\theta_u$ and $\theta_d$. As the proposed closure dispersion model is identified based on this low-frequency concentration response, its reliability may depend on the dominant frequency of the prescribed inlet concentration boundary. In particular, the proposed model is expected to provide more accurate predictions in low-frequency regimes than in high-frequency regimes.
Moreover, the analysis in \cref{sec:lead_trail} indicates that the proposed model is expected to exhibit greater accuracy and physical consistency in the leading-edge region of the temporal-spatial concentration field. This correspondence may be weakened as the characteristic transport time $L/U$ increases. Therefore, this subsection investigates the effects of both the inlet frequency $f_c$ and the characteristic transport time $L/U$ on the predictive accuracy of the proposed model.

The inlet concentration boundary is prescribed in the Fourier-series form described by \cref{eq:c_bound_fourier,eq:c_amplitude}. The fundamental concentration frequency $f_c$ is sampled over the range from $0.05\mathrm{Hz}$ to $0.5\mathrm{Hz}$. The conduit length is fixed at $1000\mathrm{mm}$. Within this computational domain, different axial positions and flow velocities are sampled to obtain a broad range of characteristic transport times, $L/U$. All other numerical settings are consistent with those described in \cref{sec:valid_pe}. The proposed model \cref{eq:patroclus_disp} and \cref{eq:conv_diff}, taken as the ground truth, are then solved numerically. A total of $200$ physical configurations are considered in this subsection, from which $1312$ cross-sectional concentration slices are sampled at different axial positions. The coefficients of determination obtained under different configurations are shown in \cref{fig:properties} (a) and (b), with the error bars indicating the corresponding STDs.

\cref{fig:properties} (a) shows that the coefficient of determination generally decreases as the fundamental concentration frequency $f_c$ increases. Although a local deviation from this overall trend is observed at $f_c=0.35\mathrm{Hz}$, it does not alter the general dependence on inlet frequency. These results confirm that the predictive accuracy of the proposed model deteriorates to some extent when the inlet concentration signal contains increasingly high-frequency components. Nevertheless, the proposed model still maintains an $R^2$ of approximately $0.8$ at the highest $f_c$ considered. Furthermore, the dominant frequency of physiologically relevant concentration pulses is expected to be substantially lower than the upper frequency considered here.

As shown in \cref{fig:properties} (b), the coefficient of determination generally decreases as the characteristic transport time $L/U$ increases. A slight increase is observed when $L/U>3\mathrm{s}$; however, its magnitude is sufficiently small that the overall decreasing trend remains unchanged. These results indicate that the predictive accuracy of the proposed model decreases as the transport time becomes longer, which is consistent with the analytical results given in \cref{sec:lead_trail}. Nevertheless, the model still achieves an $R^2$ of approximately $0.8$ at the largest $L/U$ considered. Such a transport time is substantially longer than that typically encountered in arterial contrast-agent transport, owing to the relatively high blood flow velocities and the limited length of vascular segments that can be clearly resolved and analysed in practice.

Furthermore, the dimensionless indicator

\begin{equation}
    \theta_c=(1-\theta_d)\frac{f_cL}{U}
    \label{eq:theta_c}
\end{equation}
is introduced to provide a quantitative measure of the combined effects discussed in this subsection. The product $f_cL/U$ combines the effects of the inlet concentration frequency and the characteristic transport time, while the factor $1-\theta_d$ accounts for the extent to which the proposed model departs from the classical Taylor--Aris dispersion regime. In the limit $\theta_d\rightarrow1$, the proposed model approaches the classical Taylor--Aris dispersion regime, such that the effects associated with $f_c$ and $L/U$ become negligible. Conversely, at high radial P\'eclet numbers, $\theta_d\rightarrow0$, and the model approaches the convection-dominated dispersion regime, where the effects considered in this subsection become increasingly relevant.
As shown in \cref{fig:properties} (c), the coefficient of determination generally decreases with increasing $\theta_c$. A slight increase is observed for $\theta_c>0.7$; however, its magnitude is sufficiently small that the overall decreasing trend remains unchanged. A relatively strong correlation between $R^2$ and $\theta_c$ can be observed over the considered parameter range, suggesting that $\theta_c$ provides a useful dimensionless indicator for quantifying the sensitivity of the proposed model to the inlet concentration signal and characteristic transport time.

\subsubsection{Dependence of the proposed model on the flow pulsatility}

The analysis in \cref{sec:discuss_interp} indicates that the convection-dominated transport behaviour described by the proposed dispersion model involves a redistribution of the convective flux $J_w$ into an effective convective flux $\theta_u\avg{w}\avg{C}$ and a dispersion flux $J_\mathrm{disp}$. In addition, the identification of the proposed model in \cref{sec:identification} is based on the low-frequency transfer behaviour described by \cref{eq:trans_func_steady}, which is mainly governed by the steady components of the dimensionless cross-sectionally averaged pulsatile flow velocity $\avg{w}$ and the dispersion correction factor $\gamma_w$. Consequently, the convective flux $J_w$ described by \cref{eq:c_flux_conv} is predominantly associated with the Poiseuille velocity profile corresponding to the steady flow component.
However, as the frequency of flow pulsation increases, the instantaneous velocity profile gradually departs from the classical Poiseuille solution and instead approaches the pulsatile-flow solution described by \cite{Womersley1955}. This change in the radial velocity distribution may consequently influence the flux redistribution mechanism described by the proposed dispersion model and therefore affect its predictive accuracy under strongly pulsatile flow conditions.

To investigate the effects of flow pulsatility on the proposed model, two physical indicators are considered in this subsection: the fundamental flow velocity frequency $f_u$ and the corresponding fundamental Womersley number, $\alpha$. The Womersley number provides a quantitative measure of the influence of flow pulsatility on the radial velocity profile. In general, a Womersley number can be associated with each frequency component of a pulsatile flow. In the present study, the flow velocity contains only a single frequency component. Therefore, the Womersley number considered here corresponds directly to the fundamental flow velocity frequency $f_u$ and is defined as

\begin{equation}
    \alpha = R\sqrt{\frac{2\pi f_u\rho}{\mu}},
    \label{eq:alpha}
\end{equation}
where $R$, $\mu$, and $\rho$ denote the conduit radius, blood viscosity, and mass density, respectively. The inlet concentration boundary is prescribed according to \cref{eq:c_bound_valid}. The fundamental flow velocity frequency $f_u$ is sampled over the range from $0.5\mathrm{Hz}$ to $1.5\mathrm{Hz}$. Meanwhile, the fundamental Womersley number $\alpha$ is independently sampled over the range from $1$ to $20$. This independent sampling is achieved by varying $\mu$ and $\rho$ according to \cref{eq:alpha}, allowing the effects of $f_u$ and $\alpha$ to be examined separately. All other numerical settings are consistent with those described in \cref{sec:valid_pe}.

The proposed model \cref{eq:patroclus_disp} and the full convection-diffusion equation, \cref{eq:conv_diff}, taken as the ground truth, are then solved numerically. A total of $275$ physical configurations are considered in this subsection, from which $825$ cross-sectional concentration slices are sampled at different axial positions. The coefficients of determination, $R^2$, are shown as functions of the Womersley number $\alpha$ and the fundamental flow velocity frequency $f_u$ at different axial positions in \cref{fig:properties} (d) and (e), respectively, with the error bars indicating the corresponding STDs. Three axial positions, $L=200\mathrm{mm}$, $300\mathrm{mm}$, and $400\mathrm{mm}$, are considered.
No significant correlation is observed between $R^2$ and either $\alpha$ or $f_u$ over the parameter ranges considered. These results indicate that the predictive accuracy of the proposed model is relatively insensitive to flow pulsatility, suggesting that the deviation of the instantaneous velocity profile from the classical Poiseuille profile has only a limited influence on the flux redistribution mechanism represented by the proposed model.

\section{Implication for inverse problem}

\subsection{Problem set-up}

\begin{table}
	\begin{center}
		\def~{\hphantom{0}}
		\begin{tabular}{lccc}
			Parameter&	Description&	Unit&	Value \\[3pt]
			$L$&	Axial conduit length&	$\mathrm{mm}$&	$100.0-200.0$ \\
			$R$&	Conduit radius&	$\mathrm{mm}$&	$3.0$ \\
			$U$&	Characterize flow velocity&	$\mathrm{mm/s}$&	$300.0$ \\
			$\Pen_r$&	Radial P\'eclet number&	$-$&	$10^2-10^7$ \\
			$\mu$&	Blood viscosity&	$\mathrm{mPa\cdot s}$&	$3.0$ \\
			$\rho$&	Mass density&	$\mathrm{kg/m^{3}}$&	$1060.0$ \\ 
			$f_u$&	Frequency of the sinusoidal flow velocities&	$\mathrm{Hz}$&	$1.167$ \\
			$\Delta\tau$&	Discretization interval of dimensionless time&	$-$&	$0.01$ \\
			$\Delta\xi$&	Discretization interval of dimensionless axial length&	$-$&	$0.002$ \\
			$\Delta y$&	Discretization interval of dimensionless radius&	$-$&	$0.005$ \\
		\end{tabular}
		\caption{Numerical simulation parameters for inverse problems.}
		\label{tab:num_params_inverse}
	\end{center}
\end{table}

The purpose of this section is to examine the applicability of the proposed model to flow velocity estimation from contrast-enhanced medical imaging. To this end, scalar concentration observations are numerically generated under physiologically realistic flow conditions. Based on these observations, the time-averaged cross-sectional flow velocity is estimated using both conventional velocity estimation methods based on concentration-time curves (CTCs) and their model-informed counterparts, in which the transport behaviour described by the proposed model is incorporated into the conventional estimation framework. The performance of the two approaches is then compared to assess the effect of the proposed model on the accuracy of flow velocity estimation.

\begin{figure}
	\centering
	
	\begin{subfigure}{\textwidth}
		\makebox[\linewidth][l]{(a)}
		\centering
		\includegraphics[width=\textwidth]{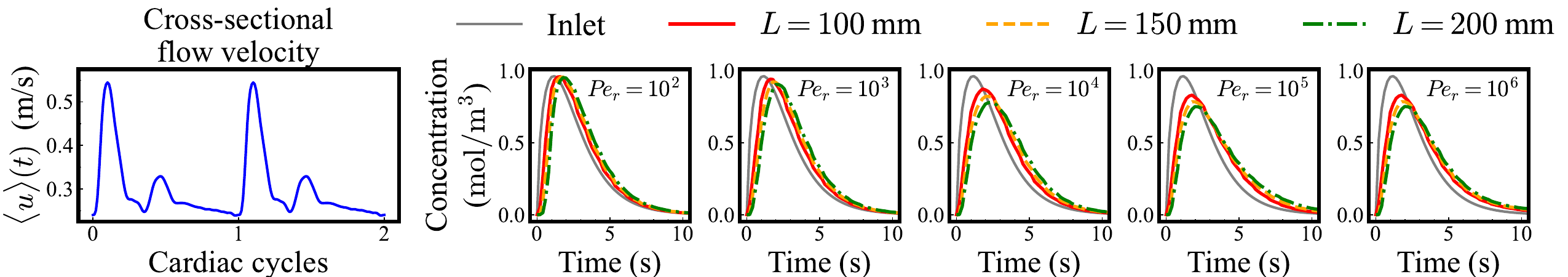}
	\end{subfigure}
	
	\vspace{0.5em}
	
	\begin{subfigure}{\textwidth}
		\makebox[\linewidth][l]{(b)}
		\centering
		\includegraphics[width=\textwidth]{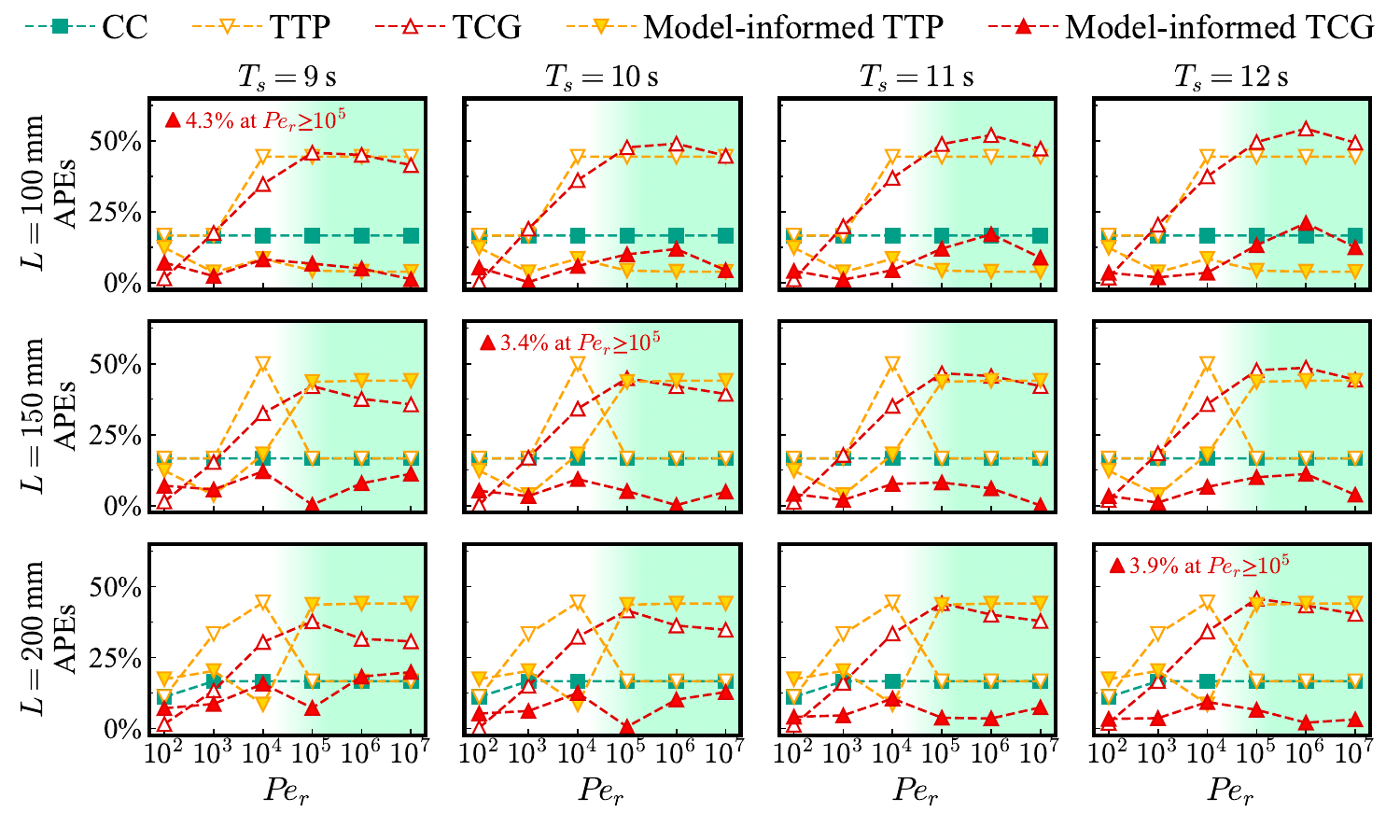}
	\end{subfigure}
	
	\caption{
		(a) Prescribed cross-sectionally averaged flow velocity and corresponding CTCs at different $\Pen_r$ and ROI lengths $L$. The concentration profiles are obtained from the forward transport simulations and subsequently used as observations for the inverse reconstruction.
        (b) APEs of the estimated flow velocity obtained using CC, TTP, TCG, model-informed TTP and model-informed TCG for different sampling times $T_s$ and ROI lengths $L$. The shaded region indicates the convection-dominated regime at high $\Pen_r$. The annotations indicate the minimum APE achieved by the model-informed TCG at $\Pen_r\geq 10^5$ for each ROI length.
	}
	\label{fig:inverse_average}
\end{figure}

The transport of a contrast agent under physiological conditions is simulated numerically by solving the dimensionless convection-diffusion equation, namely \cref{eq:conv_diff_ndim}. The flow velocity is prescribed according to the physiological waveform shown in \cref{fig:inverse_average} (a), and the inlet concentration is prescribed as a gamma-variate function described in \cref{eq:c_bound_valid}. Time integration is performed using a weighted implicit $\theta$-scheme with $\theta=0.55$. The axial convective term is approximated using a first-order upwind difference, whereas the axial and radial diffusive terms are discretized using second-order central differences. The symmetry condition at the centerline and the no-flux condition at the wall are imposed using the corresponding finite-difference boundary stencils. The numerical parameters are summarized in \cref{tab:num_params_inverse}, with all physical parameters selected within physiological ranges.

The numerical results described above are averaged over the cross-section to generate scalar concentration observations representative of those obtainable from clinical medical imaging, which are shown in \cref{fig:inverse_average} (a). These observations are then used to estimate the cross-sectional average flow velocity within the vessel using different CTC-based methods. The purpose of this section is to examine the applicability of the proposed model to CTC-based flow velocity estimation, rather than to investigate the reliability or robustness of the inverse algorithms themselves. Accordingly, only noise-free observations are considered throughout this section. The CTC-based methods considered in this section are introduced in the next subsection, together with their model-informed counterparts.

\subsection{CTC-based flow velocity estimation methods}

By selecting a specific region of interest (ROI), CTC-based velocity estimation methods infer the time-averaged cross-sectional flow velocity from the CTCs at the proximal and distal ends of the ROI. In this subsection, the CTCs at the proximal and distal ends are defined as

\begin{equation}
    \begin{cases}
        \avg{c}_p(t)=\avg{c}(t)\mid_{x=0}, \\
        \avg{c}_d(t)=\avg{c}(t)\mid_{x=L},
    \end{cases}
    \label{eq:prox_dist_ctc}
\end{equation}
respectively, where $L$ denotes the axial length of the ROI. \cref{eq:conv_diff} is solved numerically to obtain discrete samples of the CTCs in \cref{eq:prox_dist_ctc}. To mimic the temporal resolution of practical medical imaging, the CTCs used in this subsection are downsampled to $5\mathrm{Hz}$.
Five flow velocity estimation methods are considered in this subsection, including three conventional CTC-based methods and two model-informed methods. The latter incorporate the transport behaviour predicted by the proposed model \cref{eq:patroclus_disp} into the conventional estimation framework.

The conventional CTC-based methods first estimate the transit time $T_\mathrm{trans}$ of the contrast agent from the observed CTCs, and subsequently calculate the effective time-averaged transport velocity as

\begin{equation}
    U_\mathrm{eff}=\frac{L}{T_\mathrm{trans}}.
    \label{eq:u_eff_trans}
\end{equation}
These methods assume that the effective transport velocity is equivalent to the physical time-averaged flow velocity, such that

\begin{equation}
    \widehat{U}=U_\mathrm{eff},
    \label{eq:u_pred_taylor}
\end{equation}
where $\widehat{U}$ denotes the estimated flow velocity.
However, the results and physical interpretation presented in the preceding sections suggest that, under physiological high-P\'eclet-number conditions, the effective transport velocity is modulated by the coefficient $\theta_u$ described in \cref{eq:theta_u_d_explicit}. Consequently, the estimated flow velocity should instead satisfy

\begin{equation}
    \theta_u\widehat{U}=U_\mathrm{eff}.
    \label{eq:u_pred_patro_conservation}
\end{equation}
For a prescribed molecular diffusivity $D_m$ and vessel radius $R$, the model-informed estimate can therefore be obtained by solving

\begin{equation}
    \widehat{U}=\underset{\widehat{U}}{\arg\min}\left\| \Theta_u(D_m,\widehat{U},R)\widehat{U} - U_\mathrm{eff} \right\|_2.
    \label{eq:u_pred_patro}
\end{equation}

The cross-correlation (CC) algorithm \citep{Rosen1973} estimates the transit time $T_\mathrm{trans}$ by calculating the cross-correlation between the proximal and distal CTCs. Specifically, the time shift $T_\mathrm{trans}$ that maximizes the cross-correlation

\begin{equation}
    R_\mathrm{CC}=\int_0^{T_s} \avg{c}_p(t)\cdot\avg{c}_d(t-T_\mathrm{trans})dt
    \label{eq:r_cc}
\end{equation}
over the prescribed sampling interval $T_s$ is taken as the transit time between the proximal and distal ends. The corresponding flow velocity estimate $\widehat{U}$ is then obtained from \cref{eq:u_eff_trans,eq:u_pred_taylor}.
It should be noted that the CC algorithm estimates the temporal shift between the proximal and distal CTCs from their statistical correlation, rather than the physical transport time of the contrast agent. Consequently, the estimated time shift does not have the physical interpretation required by the transport correction in the proposed model. The CC algorithm is therefore used solely as a baseline method in this subsection and is not included in the comparison of the model-informed methods.

The time-to-peak (TTP) algorithm estimates the transit time of the contrast agent from the difference between the times of peak concentration at the proximal and distal ends of the ROI. Specifically, the transit time estimated by TTP algorithm is defined as

\begin{equation}
    T_\mathrm{trans}=\underset{t\in[0,T_s]}{\arg\max}\ \avg{c}_d(t)-\underset{t\in[0,T_s]}{\arg\max}\ \avg{c}_p(t)
    \label{eq:ttp_trans}
\end{equation}
where $T_s$ denotes the prescribed sampling interval. The corresponding flow velocity estimate $\widehat{U}$ is obtained from \cref{eq:u_eff_trans,eq:u_pred_taylor}.
Unlike the CC algorithm, the resulting transit time given by the TTP algorithm retains a physical interpretation as an estimate of the transport time, making it amenable to the transport correction introduced by the proposed model. Accordingly, a model-informed TTP method is obtained by replacing the conventional velocity estimate in \cref{eq:u_pred_taylor} with the corrected estimate in \cref{eq:u_pred_patro}.

The time-of-centre of gravity (TCG) algorithm \citep{Rutishauser1967} estimates the transit time of the contrast agent from the difference between the centres of gravity of the proximal and distal CTCs. For a given CTC, its centre-of-gravity time is defined as

\begin{equation}
    T_\mathrm{TCG}=\frac{\int_0^{T_s}t\avg{c}(t)dt}{\int_0^{T_s}\avg{c}(t)dt},
    \label{eq:t_tcg}
\end{equation}
where $T_s$ denotes the prescribed sampling interval. The transit time is subsequently estimated as

\begin{equation}
    T_\mathrm{trans}=\frac{\int_0^{T_s}t\avg{c}_d(t)dt}{\int_0^{T_s}\avg{c}_d(t)dt}-\frac{\int_0^{T_s}t\avg{c}_p(t)dt}{\int_0^{T_s}\avg{c}_p(t)dt}.
    \label{eq:tcg_trans}
\end{equation}
The corresponding flow velocity estimate $\widehat{U}$ is obtained from \cref{eq:u_eff_trans,eq:u_pred_taylor}. A model-informed TCG method is then constructed by replacing the conventional velocity estimate in \cref{eq:u_pred_taylor} with the corrected estimate in \cref{eq:u_pred_patro}.

\subsection{Results of flow velocity inversion}

The results of the flow velocity inversion using the CC, TTP, TCG, model-informed TTP, and model-informed TCG are illustrated in \cref{fig:inverse_average} (b). The twelve panels are arranged in three rows and four columns, with each row corresponding to a different ROI length $L$ and each column corresponding to a different sampling interval $T_s$. In each panel, the absolute percentage error (APE) of the estimated flow velocity is plotted against the radial P\'eclet number $\Pen_r$, where the APE is defined as
\begin{equation}
    \mathrm{APE}=\frac{\left|\widehat{U}-U\right|}{U} \times 100\%,
    \label{eq}
\end{equation}
with $\widehat{U}$ and $U$ denoting the estimated time-averaged flow velocity and the ground truth, respectively. The radial P\'eclet number spans from $10^2$ to $10^7$. The cyan-shaded region in \cref{fig:inverse_average} (b) indicates the range of $\Pen_r$ representative of physiological flow conditions, which is approximately above $10^5$. This range is of particular interest because it corresponds to the high-P\'eclet-number regime for which the proposed model is developed.

It is observed in \cref{fig:inverse_average} (b) that the model-informed TCG algorithm outperforms the other methods under most conditions. \cite{Wu2022} reported that the TCG algorithm was the most reliable method for estimating blood flow velocity among the algorithms considered in their study. Consistent with this finding, the present results demonstrate that the conventional TCG algorithm performs well only in the low-P\'eclet-number regime, where $\Pen_r\approx10^2$, and gradually loses accuracy as $\Pen_r$ increases for all considered conditions. In contrast, the model-informed TCG algorithm maintains good accuracy across the entire range of $\Pen_r$ considered. These results indicate that incorporating the proposed model is essential for accurate time-averaged flow velocity estimation under physiological high-P\'eclet-number conditions, while the TCG algorithm remains a promising framework for CTC-based flow velocity estimation.

The physical origin of the observed improvement can be further understood from the CTCs shown in \cref{fig:inverse_average} (a). As $\Pen_r$ increases, the trailing edge of the CTCs becomes slightly more extended. This tailing effect shifts the centre of gravity of the distal CTC towards later times, causing the conventional TCG algorithm to overestimate the transit time $T_\mathrm{trans}$ and consequently underestimate the effective transport velocity $U_\mathrm{eff}$. The resulting systematic bias becomes increasingly pronounced as $\Pen_r$ increases, leading to the deterioration in flow velocity estimation observed in \cref{fig:inverse_average} (b).

In the proposed model, this tailing behaviour is represented through the enhanced effective diffusivity in \cref{eq:deff_patroclus}. As discussed previously, at high $\Pen_r$, the enhanced dispersion in the effective 1-D model modifies the relative contributions of convection and dispersion to axial transport, such that the effective transport velocity is lower than the physical time-averaged flow velocity. This behaviour is captured by the parameter $\theta_u<1$ in the high-P\'eclet-number regime. Therefore, accounting for the reduction in the effective transport velocity through \cref{eq:u_pred_patro_conservation} provides a physically consistent correction to the velocity estimated from the transit time. This explains why the model-informed TTP and TCG methods achieve higher accuracy than their conventional counterparts, particularly under physiological high-P\'eclet-number conditions.

The results also indicate that the performance of the model-informed TCG method depends on both the ROI length $L$ and the sampling interval $T_s$. An appropriate choice of these two parameters can further improve the accuracy of flow velocity estimation in the high-P\'eclet-number regime. In \cref{fig:inverse_average} (b), for each ROI length, the value of $T_s$ that minimizes the APE is marked together with the corresponding minimum APE. Each marked value therefore represents the lowest estimation error obtained among the considered $T_s$ values for the corresponding ROI length. It is observed that the optimal $T_s$ increases with increasing ROI length. This behaviour can be attributed to the overall delay of the distal CTC as the ROI length increases, which requires a longer sampling interval to adequately cover the temporal range of the CTC and thereby obtain a reliable estimate of its centre of gravity. These results suggest a further improvement of the model-informed TCG method by adaptively selecting the ROI length and sampling interval based on the observed CTCs, potentially enabling a more accurate estimation of time-averaged flow velocity under physiological high-P\'eclet-number conditions.

\section{Conclusion}

\subsection{Main findings}

We developed an explicit dispersion model for longitudinal contrast-agent transport in arterial flows at high P\'eclet numbers. Two closure coefficients, $\theta_u$ and $\theta_d$, were introduced to account for deviations from the classical Taylor--Aris dispersion behaviour. Their values were first identified through low-frequency transfer-function matching, and their functional structures were subsequently discovered using a modified KAN, leading to a complete explicit closure model, namely \cref{eq:patroclus_disp}. Explicit expressions for $\theta_u$ and $\theta_d$ are described in \cref{eq:theta_u_d_explicit}. The proposed model was then subjected to extensive numerical validation and physical analysis. The results demonstrate that the proposed model provides a more reliable description of longitudinal transport than the classical Taylor--Aris model in the high-$\Pen_r$ regime, and that accounting for the non-classical dispersion behaviour is important for contrast-agent-based flow velocity inversion in arteries. The main findings are summarized as follows:

\begin{enumerate}[align=left]

\item[(i)] In the high-P\'eclet-number regime, axial solute transport is redistributed between the effective convection flux and the dispersive flux, resulting in a reduction of the effective convective transport velocity in the dispersion model, and giving rise to the observed convection-dominated dispersion behaviour. This reduction is characterised by $\theta_u<1$, which quantifies the deviation of the effective convection velocity from the mean flow velocity.

\item[(ii)] In the low-P\'eclet-number regime, both closure coefficients approach 1 and the proposed model reduces to the classical Taylor--Aris dispersion model. The transition between the two transport regimes is primarily characterized by the characteristic P\'eclet number $\Pen_0$.

\item[(iii)] The model accuracy is affected by the leading- and trailing-edge effects of the inlet concentration signal. The transport states represented by the proposed model are concentrated in the early-time portion of the contrast passage, where the leading edge dominates. This regime is therefore particularly relevant to the characteristic transport stage of clinical contrast-agent bolus passage, whereas the model becomes less accurate as the characteristic transport time increases.

\item[(iv)] The correction to the effective transport velocity introduced by $\theta_u$ has a non-negligible effect on CTC-based arterial blood flow velocimetry. In particular, incorporating $\theta_u$ into the model improves the reconstruction of the averaged blood flow velocity in the high-P\'eclet-number regime.

\end{enumerate}

\subsection{Limitations and future work}

Several limitations of the present study should be acknowledged. The proposed 1-D model does not include higher-order axial derivative terms, as would arise in a generalized Taylor dispersion theory \citep{Frankel1989}. Under classical Taylor--Aris dispersion, the third- and higher-order axial derivatives are associated with sufficiently small scales that their contributions to the dispersion behaviour can generally be neglected. Whether this assumption remains valid in the high-$\Pen_r$ regime is, however, not established here, because the underlying transport behaviour departs substantially from the classical Taylor--Aris-dispersion regime. The present formulation deliberately retains the lower-order effective-model structure in order to focus on the macroscopic mass-transport behaviour and, importantly, to preserve the computational efficiency required for its use as a forward model in inverse problems. Consequently, the possible contribution of higher-order axial derivatives remains an open issue for future investigation.

The accuracy of the proposed model is subject to the characteristic-time and concentration-frequency effects identified in the preceding analysis. Although the corresponding regimes are outside the physiological conditions considered in this study, they nevertheless define an intrinsic limitation of the present formulation. By retaining the structural form of a classical Taylor-dispersion model, the proposed approach represents a complex transport process through a compact cross-sectionally averaged description. This reduction enables the model to capture the dominant macroscopic transport behaviour with substantially reduced computational cost, but necessarily limits its ability to resolve more complex features of the transport process, such as memory effects and detailed boundary effects. Incorporating such phenomena would require a more elaborate model structure, potentially involving additional state variables, higher-order derivatives, or non-local transport terms. Such extensions would, however, need to be balanced against the simplicity that motivates the present effective-model formulation.

The inverse problem demonstration in this study is restricted to the estimation of the mean flow velocity. The applicability of the proposed model to transient flow velocity reconstruction has not been established. In fact, preliminary numerical experiments indicate that directly using the proposed model as the forward model within a conventional optimization framework for transient velocity reconstruction can lead to poor reconstruction results. The introduction of $\theta_u$ and $\theta_d$ substantially enriches the nonlinear dependence of the transport response on the flow parameters, thereby increasing parameter coupling and making the resulting optimization problem poorly conditioned for conventional optimization methods. Direct extension of the present framework to transient flow reconstruction therefore cannot be regarded as straightforward. A successful treatment would likely require a problem-specific inversion strategy that explicitly accounts for the nonlinear closure structure and its associated parameter coupling. The development of such a specialized inverse framework is left for future work.

\begin{bmhead}[Funding]
This work was supported by the Key Program of the National Natural Science Foundation of China (Grant No. 12632015), Dalian Municipal Guiding Program for the Life and Health Sector (no. 2025ZDJH01PT040), and the Fundamental Research Funds for the Central Universities (no. DUT25YG272).
\end{bmhead}

\begin{bmhead}[Declaration of interests]
The authors report no conflict of interest.
\end{bmhead}

\begin{appen}

\section{Parameter reformulation of \cref{eq:theta_u_d_kan}}\label{app:A}
Inspection of the optimized parameters in \cref{tab:struct_para_kan} reveals the following approximate relationships:

\begin{equation}
    \begin{cases}
        \lambda_6 \approx -\frac{1}{\log(10)}, \\
        \lambda_3\lambda_7 \approx 1, \\
        b_6 \approx 0.
    \end{cases}
	\label{eq:theta_d_param_approx}
\end{equation}
Furthermore, the following parameter transformation is introduced:

\begin{equation}
	\Pen_0 = \frac{1}{\zeta_5\e^{\lambda_7b_3+1}} \frac{U_0R_0}{D_0}.
	\label{eq:pe_0}
\end{equation}
Substituting \cref{eq:pe,eq:kan_input,eq:pe_star,eq:pe_0,eq:theta_d_param_approx} into the expression for $\theta_d^*$ in \cref{eq:theta_u_d_kan}, together with the logarithmic base-conversion identity, gives

\begin{equation}
	\theta_d^* \approx \frac{1}{1+\frac{\Pen_r}{\Pen_0}}.
    \label{eq:theta_d_approx}
\end{equation}
\cref{tab:struct_para_kan} also reveals following approximate relationships:

\begin{equation}
	\begin{cases}
	    \lambda_2 \approx -3\lambda_1, \\
        \lambda_4 \approx 1, \\
        b_7 \approx 1.
	\end{cases}
    \label{eq:theta_u_param_approx}
\end{equation}
The first relation in \cref{eq:theta_u_param_approx} suggests that the exponents associated with $U^*$ and $R^*$ approximately satisfy a fixed ratio. This observation motivates the introduction of \cref{eq:beta}
Substituting \cref{eq:kan_input,eq:theta_u_param_approx,eq:beta} together with parameter transformations

\begin{equation}
    \begin{cases}
        \Pen_1 = \frac{1}{{\zeta_2}^{\frac{1}{\lambda_4}}} \frac{U_0R_0}{\e D_0}, \\
        \beta_0 = \zeta_4^{\frac{1}{\lambda_1\lambda_5}} \e^{\frac{b_1+b_2}{\lambda_1}-2} \frac{{R_0}^3}{U_0}
    \end{cases}
    \label{eq:theta_u_param_trans_1}
\end{equation}
into the expression for $\theta_u^*$ in \cref{eq:theta_u_d_kan}, gives

\begin{equation}
    \theta_u^* \approx \zeta_6 \sigma\left[
        -\frac{\zeta_1\lambda_8}{1+\frac{\Pen_r}{\Pen_1}} - \frac{\zeta_3\lambda_8}{1+\left(\frac{\beta}{\beta_0}\right)^{-\lambda_1\lambda_5}} - \lambda_8(b_4 + b_5) - \log(\zeta_7)
    \right] + 1,
    \label{eq:theta_u_hill}
\end{equation}
where $\sigma(\cdot)$ denotes the sigmoid function defined by

\begin{equation}
    \sigma(x)=\frac{1}{1+\exp(-x)}.
    \label{eq:sigmoid}
\end{equation}
To further improve the compactness of the closure structure, the number of independent parameters is reduced wherever possible. The first term in the sigmoid argument of \cref{eq:theta_u_hill} shares the same functional dependence on the radial P\'eclet number as that of \cref{eq:theta_d_approx}. Since the optimized values of $Pe_0$ and $Pe_1$ are of the same order of magnitude, this term is approximated by $\theta_d^*$, namely

\begin{equation}
    \frac{1}{1+\frac{\Pen_r}{\Pen_1}} \approx \theta^*_d.
    \label{eq:pe_1_approx}
\end{equation}
Substituting \cref{eq:pe_1_approx} together with parameter transformations

\begin{equation}
    \begin{cases}
        \zeta_u = -\zeta_6, \\
        \zeta_d = -\zeta_1\lambda_8, \\
        \zeta_{\beta} = -\zeta_3\lambda_8, \\
        \lambda_{\beta} = -\lambda_1\lambda_5, \\
        b_u = -\lambda_8(b_4 + b_5) - \log(\zeta_7)
    \end{cases}
    \label{eq:theta_u_param_trans_2}
\end{equation}
into \cref{eq:theta_u_hill} yields

\begin{equation}
    \theta_u^* \approx 1 - \zeta_u \sigma\left[
        \zeta_d \theta_d^* + \frac{\zeta_{\beta}}{1+\left(\frac{\beta}{\beta_0}\right)^{\lambda_{\beta}}} + b_u
    \right].
    \label{eq:theta_u_approx}
\end{equation}

\end{appen}\clearpage

\bibliographystyle{jfm}
\bibliography{ref_patroclus}
	
\end{document}